\documentclass[preprint,12pt]{elsarticle}

\journal{Computer Physics Communications}

\usepackage{graphicx}
\usepackage{dcolumn}
\usepackage{bm}
\usepackage{natbib}
\usepackage{caption}
\usepackage{subcaption}
\usepackage{float}
\usepackage{amsmath,amssymb}
\usepackage{amsmath}
\usepackage{braket}
\usepackage{xcolor}
\usepackage{verbatim}
 
\usepackage[colorinlistoftodos, textsize=small]{todonotes}

\newcommand{\subscript}[1]{\ensuremath{_{\textrm{#1}}}}

\def\forceat{{\bm{F}_{at}}}
\def\stresstens{{\sigma_{\lambda\mu}}}
\def\bise{{Bi\subscript{2}Se\subscript{3}}}

\begin{document}

\begin{frontmatter}

\title{GPGPU Acceleration of All-Electron Electronic Structure Theory Using Localized Numeric Atom-Centered Basis Functions}

\author[DukeMEMS]{William P. Huhn\fnref{willaddress}}
\author[DukeMEMS]{Bj\"{o}rn Lange\fnref{bjoernaddress}}
\author[DukeMEMS]{Victor Wen-zhe Yu}
\author[ONLCNMS]{Mina Yoon}
\author[DukeMEMS,DukeChem]{Volker Blum\corref{correspondingauthor}}

\address[DukeMEMS]{Department of Mechanical Engineering and Materials Science, Duke University, Durham, NC 27708, USA}
\address[ONLCNMS]{Center for Nanophase Materials Sciences, Oak Ridge National Laboratory, Oak Ridge, TN 37830, USA}
\address[DukeChem]{Department of Chemistry, Duke University, Durham, NC 27708, USA}

\cortext[correspondingauthor]{Corresponding author}
\ead{volker.blum@duke.edu}

\fntext[willaddress]{Present address: Argonne Leadership Computing Facility, Argonne National Laboratory, Argonne, IL 60439, USA}
\fntext[bjoernaddress]{Present address: DACS Laboratories GmbH, Erkrath, Nordrhein-Westfalen 40699, Germany}
\begin{abstract}
We present an implementation of all-electron density-functional theory for massively parallel GPGPU-based platforms, using localized atom-centered basis functions and real-space integration grids. Special attention is paid to domain decomposition of the problem on non-uniform grids, which enables compute- and memory-parallel execution across thousands of nodes for real-space operations, e.g. the update of the electron density, the integration of the real-space Hamiltonian matrix, and calculation of Pulay forces. To assess the performance of our GPGPU implementation, we performed benchmarks on three different architectures using a 103-material test set. We find that operations which rely on dense serial linear algebra show dramatic speedups from GPGPU acceleration: in particular, SCF iterations including force and stress calculations exhibit speedups ranging from 4.5 to 6.6. For the architectures and problem types investigated here, this translates to an expected overall speedup between 3-4 for the entire calculation (including non-GPU accelerated parts), for problems featuring several tens to hundreds of atoms. Additional calculations for a 375-atom Bi$_2$Se$_3$ bilayer show that the present GPGPU strategy scales for large-scale distributed-parallel simulations.
\end{abstract}


\end{frontmatter}


\section{Introduction}

Kohn-Sham density-functional theory (KS-DFT)~\cite{PHohenberg64,WKohn65} is the primary tool for computational materials prediction across a wide range of areas in science and engineering~\cite{KBurke12,PsiK14,ADBecke14,AJain16}.  One class of KS-DFT codes uses spatially localized atom-centered basis sets, e.g. Gaussian orbitals~\cite{MJFrisch84,THDunning89,ASzabo96,AKWilson96,FWeigend05,MValiev10,JHutter14,FFurche14,YShao15,Gaussian16,SRJensen17}, Slater orbitals~\cite{JCSlater30,GteVelde01,EvanLenthe03}, and numeric atom-centered orbitals (NAOs).~\cite{FWAverill73,AZunger77,BDelley82,OFSankey89,BDelley90,APHorsfield97,KKoepernik99,GteVelde01,JMSoler02,TOzaki05,Blum09,IYZhang13} For these basis sets, local operations such as Hamiltonian and overlap matrix integrations, updates of the electron density and its gradients, or parts of force and stress tensor computations, can be carried out on real-space grids. Due to the locality of the basis functions and of the Kohn-Sham potential, these operations can be implemented such that the computational timings scale linearly in number of atoms.~\cite{JMPerezJorda95,REStratmann96,CFonsecaGuerra98,GEScuseria99,GteVelde01,JMSoler02,openMXManual,Havu09} Accordingly, in the system size range for which real-space operations dominate the cost -- typically small-to-mid-sized calculations comprising several tens to hundreds of atoms -- this translates into approximately linear-scaling overall execution times as well.  For larger-scale calculations, most KS-DFT codes have a formal default $O(N_{atoms}^{3})$ scaling due to the reliance on eigenvalue solvers to directly solve the Kohn-Sham eigenvalue problem for all occupied electron states, known as the ``cubic wall'' of standard KS-DFT. However, much of the practical science addressed by KS-DFT occurs below this limiting regime and instead in the range where operations with lower scaling exponents still account for the majority of the cost.

An increasingly available paradigm in high-performance computing (HPC) are general purpose graphics processing unit (GPGPU)-accelerated architectures, in which each computational node contains one or more GPGPU accelerators working in tandem with the traditional CPUs employed in HPC applications.  GPGPUs are well suited for evaluation of highly vectorizable, compute-intensive algorithms due to their unique massively-parallel design, prompting early work to demonstrate matrix multiplications~\cite{ESLarsen01} and FFTs~\cite{KMoreland03} on commodity hardware.  The feasibility of GPU-accelerated electronic structure calculations was demonstrated in works by Ufimtsev and Mart{\'i}nez~\cite{ISUfimtsev08,ISUfimtsev09_scf,ISUfimtsev09_scf_correction,ISUfimtsev09_grads} which would form the basis for the TeraChem electronic structure package~\cite{NLuehr16}, as well as by Yasuda using the Gaussian package~\cite{KYasuda07,KYasuda08}.  Since then, a number of electronic structure packages have incorporated GPU acceleration, including ABINIT~\cite{XGonze16}, ADF~\cite{HvanSchoot16}, BigDFT~\cite{LGenovese09,LGenovese16}, CP2K~\cite{OSchutt16}, GPAW~\cite{SHakala13,JYan13,SHakala16}, LS3DF~\cite{WJia17}, octopus~\cite{XAndrade12,Xandrade13,XAndrade16}, ONETEP~\cite{KWilkinson13}, PEtot (now PWmat)~\cite{LWang11,WJia13_1,WJia13_2,PWmat}, Q-Chem~\cite{LVogt08,ROlivaresAmaya10,YShao15}, Quantum ESPRESSO~\cite{FSpiga11,JRomero18}, RMG~\cite{SMoore12,RMG},  VASP~\cite{SMaintz11,MHutchinson12,MHacene12,MHutchinson16}, and FHI-aims (this work). Development cost can be alleviated by using drop-in GPU-accelerated libraries such as cuBLAS~\cite{cuBLAS}, cuFFT~\cite{cuFFT}, Thrust~\cite{Thrust}, ELPA~\cite{ELPA,PKus19_book,PKus19_article}, and MAGMA~\cite{STomov10_article,STomov10_conference,JDongarra14} for general mathematical operations.  Nevertheless, GPU acceleration is generally not a trivial task due to the need to target low-level algorithms, which constitute the majority of the computational workload, specific to a given software package and port them to a new architecture.  However, the raw computational output afforded by GPGPUs can be immense.

In this paper, we describe a careful GPGPU adaption and analysis of several of the dominant real-space operations for KS-DFT in the NAO-based full-potential, all-electron electronic structure code FHI-aims.~\cite{Blum09,Havu09,FKnuth15} FHI-aims is a general-purpose electronic structure simulation code, offering proven scalability to thousands of atoms and on very large, conventional distributed-parallel high-performance computers.~\cite{LNemec13,ELPA,SVLevchenko15,VWZYu18} The code achieves benchmark-quality accuracy for semi-local~\cite{lejaeghere2016reproducibility,SRJensen17}, hybrid~\cite{XRen12,Ihrig2015,SRJensen17}, and many-body perturbative~\cite{XRen12,GW100} levels of theory. The specific implementation described here helps unlock the potential of GPGPU architectures for a broad range of production simulations using semilocal DFT, including generalized gradient approximation (GGA) and meta-GGA exchange-correlation functionals. We specifically target Hamilton and overlap matrix integrations, updates of the density and its gradients, and the computation of forces and stress tensor components. 

In FHI-aims, linear scaling in the real-space operations covered in this paper is achieved using a real-space domain decomposition (RSDD) algorithm~\cite{Havu09}, wherein the set of all real-space integration points is subdivided into compact sets of points known as ``batches''.  As shown below, this RSDD algorithm is naturally parallel in both memory and workload and thus allows for an efficient load-balanced, distributed-parallel GPGPU-assisted implementation.
The problem sizes for each individual batch of points are naturally suited for GPGPU acceleration, allowing for the usage of drop-in GPGPU-accelerated libraries with minimal code restructuring necessary for the Hamiltonian matrix integrals, update of density and density gradients, and total energy gradients. The electrostatic potential calculation uses a multipole summation algorithm differing from the RSDD algorithm presented here and thus was not GPGPU-accelerated for this work. Likewise, the exact-exchange operator of hybrid DFT and eigenvalue solutions that can be accessed through libraries such as ELPA~\cite{ELPA,PKus19_book,PKus19_article} and MAGMA~\cite{STomov10_article,STomov10_conference,JDongarra14} are not targeted by this work. 

This paper is organized as follows. First, we outline the fundamental equations entering into real-space Kohn-Sham density functional theory. Next, we discuss design principles for numerical algorithms which facilitate optimal usage of GPGPU resources, the RSDD algorithms used for evaluating integrals with FHI-aims and their adaption for execution on GPGPU resources.  Finally, timing and scaling benchmarks for four different GPGPU-accelerated architectures and two sets of materials demonstrate GPGPU speedups for real-space integrations across a broad class of architectures and systems encountered in electronic structure based materials simulations.

\section{Background}

Within the Born-Oppenheimer approximation, the objective of KS-DFT is
to determine the electron density $n(\bm{r})$ for a given set of fixed
nuclear position $\{\bm{R}_{at}\}$ of each atom (the ``system
geometry''). The ground-state total energy $E_{tot}[n]$
and other observables are then evaluated as functionals of $n(\bm{r})$. Throughout this section, we assume a non-spin-polarized system for simplicity. However, the extension to collinearly spin-polarized systems is straightforward, as is the extension to spin-orbit coupled and/or non-collinear spin systems,~\cite{WPHuhn17} since the underlying matrix integrals, density and gradient updates follow exactly analogous formulae. The density is determined by solving an effective single-particle Hamiltonian $\hat{h}_{KS}$. In scalar-relativistic form, 
\begin{equation}
\begin{split}
   \hat{h}_{KS} = \hat{t}_{s} + \hat{v}_{ext} + \hat{v}_{H}[n] + \hat{v}_{xc}[n]
  \label{eq:Hamiltonian}
\end{split}
\end{equation}
where $\hat{t}_{s}$ is the (scalar-relativistic) kinetic energy,
$\hat{v}_{ext}$ is the external potential, $\hat{v}_{H}[n]$ is the
Hartree potential of the electrons, and $\hat{v}_{xc}[n]$ is the
exchange-correlation potential. Evidently, the density $n(\bm{r})$
depends on the Hamiltonian $\hat{h}_{KS}$ and the Hamiltonian depends on the
density. Finding a stationary density that yields the particular
Hamiltonian that generates this density is a
non-linear optimization problem and is cast as a ``self-consistent
field'' (SCF) cycle. 

\begin{figure}
  \includegraphics[width=\textwidth]{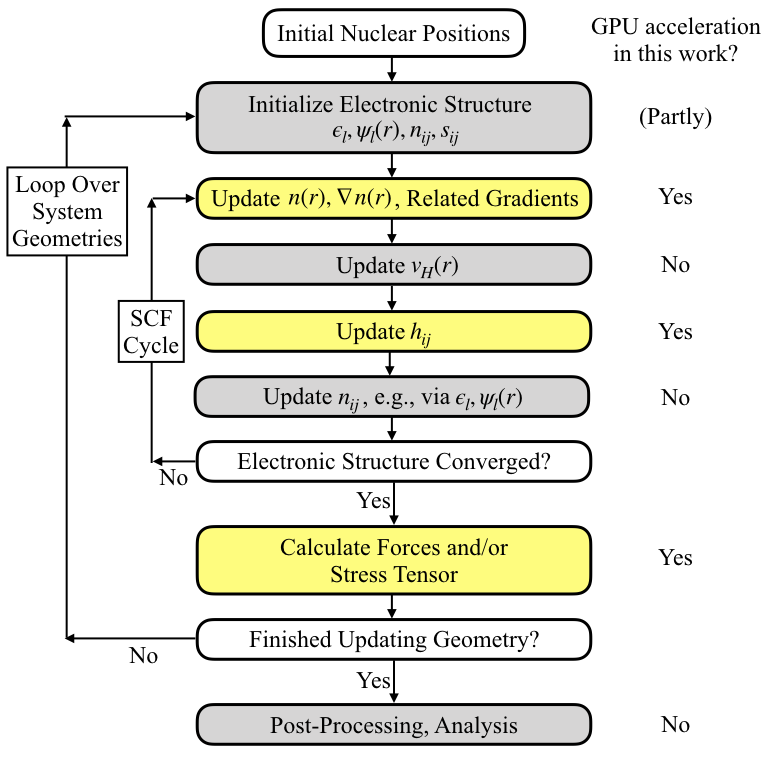}
  \caption{Program flow for a typical electronic structure calculation
    involving geometry relaxation or molecular dynamics. Shaded boxes
    indicate steps contributing to the actual computational
    workload. Yellow shading indicates steps that are subject to
    real-space GPU acceleration in this work, whereas gray shading
    indicates steps that are GPU-accelerated only partly, not at all,
    or that are handled by separate software
    components~\cite{TAuckenthaler11,ELPA,PKus19_article,PKus19_book,VWZYu18} outside the 
    scope of this work.} 
  \label{fig:SCF_Cycle}
\end{figure}

The complete flow of a typical electronic
structure calculation is shown in Figure~\ref{fig:SCF_Cycle}, which also indicates the specific operations that are GPU-accelerated in this work. For a given set of initial nuclear positions, the eigenfunctions $\{\psi_{l}\}$ (with eigenvalues $\epsilon_l$) and/or the stationary density of the Hamiltonian in Eq. (\ref{eq:Hamiltonian}) are expressed in terms of a finite set of basis functions $\{\varphi_{i}\}$. For a non-periodic system, this expansion has the form
\begin{equation}
  \psi_{l}(\bm{r}) = \sum_{i=1}^{N_{b}}c_{il}\varphi_{i}(\bm{r}).
  \label{eq:EVNonPeriodic}
\end{equation}
$N_{b}$ is the size of the basis set and $c_{il}$ is the coefficient of the $l^{\mathrm{th}}$ eigenvector for the $i^{\mathrm{th}}$ basis function.  
The density can then be computed as 
\begin{equation}
  n(\bm{r}) = \sum_{ij} \varphi^{*}_{i}(\bm{r})n_{ij}\varphi_{j}(\bm{r}),
  \label{eq:Density}
\end{equation}
where $n_{ij}$ is the density matrix, defined as
\begin{equation}
  n_{ij} = \sum_{l}f_{l}c^{*}_{il}c_{jl} ,
  \label{eq:DM_from_EV}
\end{equation}
where $f_{l}$ is the occupation number of orbital $\psi_{l}$.  If $n_{ij}$ is already known, Eq. (\ref{eq:DM_from_EV}) may be implemented in an $O(N)$ approach for localized basis elements.

Initial guesses for the electronic structure (Figure~\ref{fig:SCF_Cycle}) can either be given in terms of
$\psi_{l}(\bm{r})$ or in terms of $n_{ij}$. In FHI-aims, these initial quantities are produced by solving the Hamiltonian in Eq. (\ref{eq:Hamiltonian}) for the potential of a sum of overlapping free-atom densities.  The density gradient $\nabla n$ is required for GGA calculations, where the Hessian $\nabla^{2} n$ dependence in the exchange-correlation potential energy may be converted to a $\nabla n$ dependence via integration by parts (Eq. (30) of~\cite{Blum09}) The Hessian is required for any force computation, or when evaluation of the explicit Kohn-Sham potential for GGAs is needed for any other reason.

In this work, we use localized atom-centered basis functions, which are generally non-orthonormal to
one another and yield a non-trivial overlap matrix 
\begin{equation}
  s_{ij} = \int d\bm{r} [\varphi_{i}(\bm{r})\varphi_{j}(\bm{r})].
  \label{eq:OverlapNoPeriodic}
\end{equation}

For periodic systems, we discretize the eigenvectors in terms of Bloch
orbitals $\{\chi_{i\bm{k}}\}$ that are extended throughout the entire crystal:
\begin{equation}
  \psi_{l\bm{k}}(\bm{r}) = \sum_{i=1}^{N_{b}}c_{il}(\bm{k})\chi_{i\bm{k}}(\bm{r}).
  \label{eq:EVPeriodic}
\end{equation}
$\bm{k}$ is the crystal momentum quantum number and $N_b$ is the number of basis functions associated with a single unit cell. Each $\chi_{i\bm{k}}$ is associated with a particular localized basis function ${\varphi_{i}}(\bm{r})$ and its periodic images in all other unit cells (labelled by lattice translation vectors $\bm{T}$) via the transformation
\begin{equation}\label{Eq:Bloch-basis}
  \chi_{i\bm{k}}(\bm{r}) = \sum_{\bm{T}}e^{i\bm{k}\cdot\bm{T}}\varphi_{i}(\bm{r}-\bm{T}).
\end{equation}
As a result, $\chi_{i\bm{k}}(\bm{r})$ is normalized per unit cell. The
relevant overlap matrix is then
\begin{equation}
  s_{ij}(\bm{k}) = \int_{\mathrm{unit~cell}} d\bm{r} [\chi^{*}_{i\bm{k}}(\bm{r})\chi_{j\bm{k}}(\bm{r})].
  \label{eq:OverlapPeriodic}
\end{equation}
Importantly, for spatially localized real-space basis functions
$\{\varphi_{i}(\bm{r})\}$, and using Eqs. (\ref{eq:EVPeriodic}) and
  (\ref{Eq:Bloch-basis}), the density matrix can also be expressed 
  in terms of the real-space basis functions, just like in
  Eq. (\ref{eq:Density}). The difference is that the sum in Eq. 
  (\ref{eq:Density}) now runs over real-space basis functions and
  their images both inside and outside a given unit
  cell. More precisely, since only grid points $\bm{r}$ inside a single
  cell (say, the cell at $\bm{T}$=(0,0,0)) need be considered, $i$ and
  $j$ in Eq. (\ref{eq:Density}) run from 1, ..., $N_\mathrm{rs}$, where $N_\mathrm{rs}$ is the
  number of all localized real-space basis functions in the crystal
  that are non-zero somewhere inside the volume of the unit cell at
  $\bm{T}$=(0,0,0).

As mentioned above and as indicated in Figure~\ref{fig:SCF_Cycle}, the
density computation Eq.~(\ref{eq:Density}), which has the form of
matrix multiplications, is GPU-accelerated in this work. Additionally,
the calculation of density gradients $\nabla n(\bm{r})$ is necessary
for almost any current density functional approximations, i.e., GGA and
beyond. Finally, the Hessian matrix of the density,
$\partial_{x_i} \partial_{x_j} n(\bm{r})$ ($i,j$=1,...,3), is needed
at least for total energy gradient calculations (forces and
stresses). For a given density matrix, these gradients are
straightforward extensions of Eq. (\ref{eq:Density}), but their
evaluation can still
contribute significantly to the overall computational cost: $\nabla
n(\bm{r})$ has three components, $\partial_{x_i} \partial_{x_j}
n(\bm{r})$ has six components, and each component has the same cost as
the evaluation of $n(\bm{r})$ itself.  

As a next step, the Hartree potential $\hat{v}_{H}[n]$ in Eq.
(\ref{eq:Hamiltonian}) is a local operator of the form 
\begin{equation}
  \hat{v}_{H}[n](\bm{r}) = \int d\bm{r'} \frac{n(\bm{r'})}{|\bm{r}-\bm{r'}|} \, .
  \label{eq:HartreePotential}
\end{equation}
In practice, $\hat{v}_{ext}$ (which includes the potential due to the
nuclei) and $\hat{v}_{H}[n]$ (the averaged electrostatic potential due
to the electron density) are treated together in order to evaluate an
overall charge-neutral system. As mentioned above, this full
electrostatic potential may be computed via a multipole summation
scheme~\cite{Blum09} and is not subject to GPGPU acceleration in this
work. In the system range investigated here, it does not dominate the
computational cost. 

In non-periodic system geometries and for real-valued local basis
functions $\varphi_{i}(\bm{r})$, the Hamiltonian is naturally
expressed in terms of matrix elements $h_{ij}$ with an integral form
of 
\begin{equation}
  h_{ij} = \int d\bm{r}[\varphi_{i}(\bm{r})\hat{h}_{KS}\varphi_{j}(\bm{r})].
  \label{eq:RealSpaceHamNonPeriodic}
\end{equation}
In periodic boundary conditions, 
\begin{equation}
 h_{ij}(\bm{k}) = \int_{\mathrm{unit~cell}} d\bm{r}[\chi^{*}_{i\bm{k}}(\bm{r})\hat{h}_{KS}\chi_{j\bm{k}}(\bm{r})] \, .
  \label{eq:HamPeriodic}
\end{equation}
In the periodic case, each matrix element $h_{ij}(\bm{k})$ can still be summed up from local real-space integrals (the equivalent of Eq. (\ref{eq:RealSpaceHamNonPeriodic})) of the following type:
\begin{equation}
  h_{ij}^{\mathrm{uc}} = \int_\mathrm{unit~cell} d\bm{r} [\varphi_{i}(\bm{r})\hat{h}_{KS}\varphi_{j}(\bm{r})] .
  \label{eq:RealSpaceHamPeriodic}
\end{equation}
However, in contrast to the non-periodic case, the running indices $i$
and $j$ for $h_{ij}^{\mathrm{uc}}$ in
Eq.~(\ref{eq:RealSpaceHamPeriodic}) are not restricted to 1, ...,
$N_b$ (the number of basis functions associated with a single unit
cell) but instead run over 1, ..., $N_\mathrm{rs}$, the number of localized real-space basis functions in the crystal that are non-zero somewhere inside the volume of the unit cell at $\bm{T}$=(0,0,0). The local integrals $h_{ij}^{\mathrm{uc}}$ can be computed using the exact same real-space integration code as their non-periodic equivalents. The only difference is that each integration point $\bm{r}$ is mapped back to its periodic image within the unit cell at $\bm{T}$=(0,0,0), so that the local integrals in Eq. (\ref{eq:RealSpaceHamPeriodic}) are restricted to that unit cell. The full Bloch integrals $h_{ij}(\bm{k})$ can then be constructed by inserting Eq.~(\ref{Eq:Bloch-basis}) into Eq.~(\ref{eq:HamPeriodic}). The expression for $h_{ij}(\bm{k})$ becomes a sum over individual real-space integrals of form $h_{ij}^{\mathrm{uc}}$, multiplied by Bloch phase factors $e^{i\bm{k}\cdot(\bm{T}_i-\bm{T}_j)}$ that connect the unit cells where the real-space basis functions $i$ and $j$ are centered. The overlap matrix elements $s_{ij}(\bm{k})$ can be constructed analogously.
To unify the notation between periodic and non-periodic systems, we
will refer to both Eqs. (\ref{eq:RealSpaceHamNonPeriodic}) and
(\ref{eq:RealSpaceHamPeriodic}) as the ``real-space Hamiltonian matrix''
$h^{uc}_{ij}$ for the remainder of this paper. As shown in
Figure~(\ref{fig:SCF_Cycle}), its computation is GPU-accelerated
in this work.

Given the updated KS-DFT Hamiltonian, the Hamiltonian's approximate
eigenvectors $\psi_{l\bm{k}}(\bm{r})$ and eigenvectors
$\epsilon_{l\bm{k}}$ can now be found by solving  
\begin{equation}
  \sum_{j}h_{ij}(\bm{k})c_{jl}(\bm{k}) = \epsilon_{l\bm{k}}\sum_{j}s_{ij}(\bm{k})c_{jl}(\bm{k})
  \label{eq:KS-EV-Prob}
\end{equation}
(for non-periodic systems, the $\bm{k}$ index may just be omitted). A
new density matrix $n_{ij}$ can thus be found. The eigenvector-based
approach Eq.~(\ref{eq:KS-EV-Prob}) scales computationally at least as
$O(N_{e}^{3})$, where $N_e$ is the number of electrons in the
system. Alternatively, the KS eigenvalue problem may be 
circumvented entirely by density-matrix-based
solvers,~\cite{SGoedecker99,on_bowler_2012,pexsi_lin_2013,VWZYu18} often with
reduced-dimensional scaling. In FHI-aims, the
solution of the KS eigenvalue equation in matrix form is performed via
the dense generalized eigensolver library 
ELPA~\cite{TAuckenthaler11,ELPA,PKus19_article,PKus19_book} or circumvented by other solvers
interfaced through ELSI~\cite{VWZYu18}, an open-source library which
provides an interface layer between KS-DFT codes and methods that
solve or circumvent the Kohn-Sham eigenvalue problem in
density-functional theory. The GPU acceleration or circumvention of
Eq.~(\ref{eq:KS-EV-Prob}) is thus not the topic of this paper; however, options exist in the form of the open-source, GPU-accelerated ELPA and MAGMA eigensolver libraries that are benchmarked elsewhere in the literature.~\cite{PKus19_book,PKus19_article,STomov10_article,STomov10_conference,JDongarra14}

After the SCF cycle for a given geometry is complete, electronic
structure calculations may continue by changing the geometry of a 
material as the calculation progresses, e.g. to find a local minimum-energy
geometry at the Born-Oppenheimer surface, or for molecular dynamics.  In these cases, the
position of atomic nuclei, $\bm{R}_{at}$, are updated by calculating the forces
on each atom, 
\begin{equation}
  \forceat = -\frac{\partial}{\partial \bm{R}_{at}}E_{tot} \, ,
  \label{eq:forces}
\end{equation}
as a function of the electronic structure after the SCF cycle has
converged. Here, $E_{tot}$ is the Born-Oppenheimer total energy of
the material and the subscript $at$ labels different atoms for convenience.  For localized basis elements, the computationally
expensive portion of Eq. (\ref{eq:forces}) is the Pulay forces,
which on the scalar-relativistic -- here, the atomic zero-order regular approximation (atomic ZORA; cf. Eqs. (55) and (56) in ~\cite{Blum09}) -- and GGA level have the
form
\begin{equation}
    \bm{F}^{P}_{at} = \bm{F}^{P,local}_{at} + \bm{F}^{GGA}_{at} + \bm{F}^{at.ZORA}_{at}
\end{equation}
where
\begin{equation}
\begin{split}
    \bm{F}^{P,local}_{at} = &-2 \sum_{ij} \int_\mathrm{unit~cell} d\bm{r}[[\nabla_{at}\varphi_{i}(\bm{r})]n_{ij}\hat{h}_{KS}\varphi_{j}(\bm{r})] \\
                      &+ 2 \sum_{ij} \int_\mathrm{unit~cell} d\bm{r} [[\nabla_{at}\varphi_{i}(\bm{r})]q_{ij}\varphi_{j}(\bm{r})],
\end{split}
\end{equation}
is the local (density-only) parts of the Pulay forces, 
\begin{equation}
  q_{ij} = \sum_{l}f_{l}\sum_{\bm{k}}\epsilon_{l\bm{k}}c^{*}_{il}(\bm{k})c_{jl}(\bm{k}),
  \label{eq:E_Weighted_DM}
\end{equation}
is the $q_{ij}$ energy-weighted density matrix,
\begin{equation}
\begin{split}
    \bm{F}^{GGA}_{at} = &-4 \sum_{ij} \int d\bm{r}[[\nabla_{at}\varphi_{i}(\bm{r})]n_{ij}[\frac{\partial f_{xc}}{\partial | \nabla n|^{2}} \nabla \varphi_{j}(\bm{r})\cdot \nabla n(\bm{r})]] \\
                        &-4 \sum_{ij} \int d\bm{r}[\varphi_{i}(\bm{r})n_{ij}[\frac{\partial f_{xc}}{\partial | \nabla n|^{2}} \nabla_{at}\nabla \varphi_{j}(\bm{r})\cdot \nabla n(\bm{r})]] \\
\end{split}
\end{equation}
is the correction to the Pulay forces arising from explicit density gradients in GGA, and
\begin{equation}
    \bm{F}^{at.ZORA}_{at} = -\sum_{ij} \int d\bm{r}[[\nabla_{at}\hat{t}_{\mathrm{at.ZORA}}\varphi_{i}(\bm{r})]n_{ij}\varphi_{j}(\bm{r})]
\end{equation}
is the correction to the Pulay forces arising from atomic ZORA.

For periodic calculations, one may also calculate
overall stresses on the computational cell by calculating the stress
tensor  
\begin{equation}
  \stresstens = \frac{1}{V}\frac{\partial E_{tot}}{\partial \epsilon_{\lambda\mu}}\bigg\rvert_{\epsilon=0}
  \label{eq:stress_tensor}
\end{equation}
after the SCF cycle has converged. Here, $\epsilon_{\lambda\mu}$ is
the symmetric, infinitesimal strain tensor and $V$ is the volume of
the computational cell. A full account of the terms that make up the stress tensor in an atom-centered basis set is given in~\cite{FKnuth15}.

As mentioned above, the computation of the
gradients Eqs. (\ref{eq:forces}) and (\ref{eq:stress_tensor}) using
atom-centered basis functions for semilocal density functionals can
include the numerically expensive evaluation of higher density
derivatives such as the density Hessian and similar quantities
evaluated on a real-space grid. When needed, the computational cost of
Eqs. (\ref{eq:forces}) and (\ref{eq:stress_tensor}) can therefore
amount to a large fraction of the overall time required to execute the
type of computation shown in 
Figure~\ref{fig:SCF_Cycle}. 
The corresponding steps are thus also 
GPU accelerated in this work, using the same real-space grid
techniques as for the density and the Hamilton matrices, outlined below.

\section{Implementation}

\subsection{Design Principles for GPGPU-Accelerated Code}

There are three major design priorities entering into GPGPU acceleration of an algorithm:
\begin{itemize}
\item The workload offloaded to the GPGPU should be highly vectorizable,
\item Thread-divergent branching statements (e.g. ``if'' statements that can lead to 
different conditional states for different threads) should be avoided in the workload offloaded to the GPGPU, and
\item Communication between the CPU and GPGPU over relatively slow buses should be minimized.
\end{itemize}

Sections of an algorithm which are not easily vectorizable and/or contain a large number of thread-divergent branching statements should be performed by the CPU(s), and sections of an algorithm that are easily vectorizable should be offloaded to the GPGPU.  Ideally, GPGPU and CPU workloads should be carried out in parallel in an asynchronous fashion if that is possible.  There is an additional design priority implicit in this scheme:  the problem size for the workload offloaded to the GPGPU must fit into the GPGPU's onboard memory.

\subsection{Computational Choices in FHI-aims}

\subsubsection{Real-Space Integration Grids}

\begin{figure}
  \includegraphics[width=\textwidth]{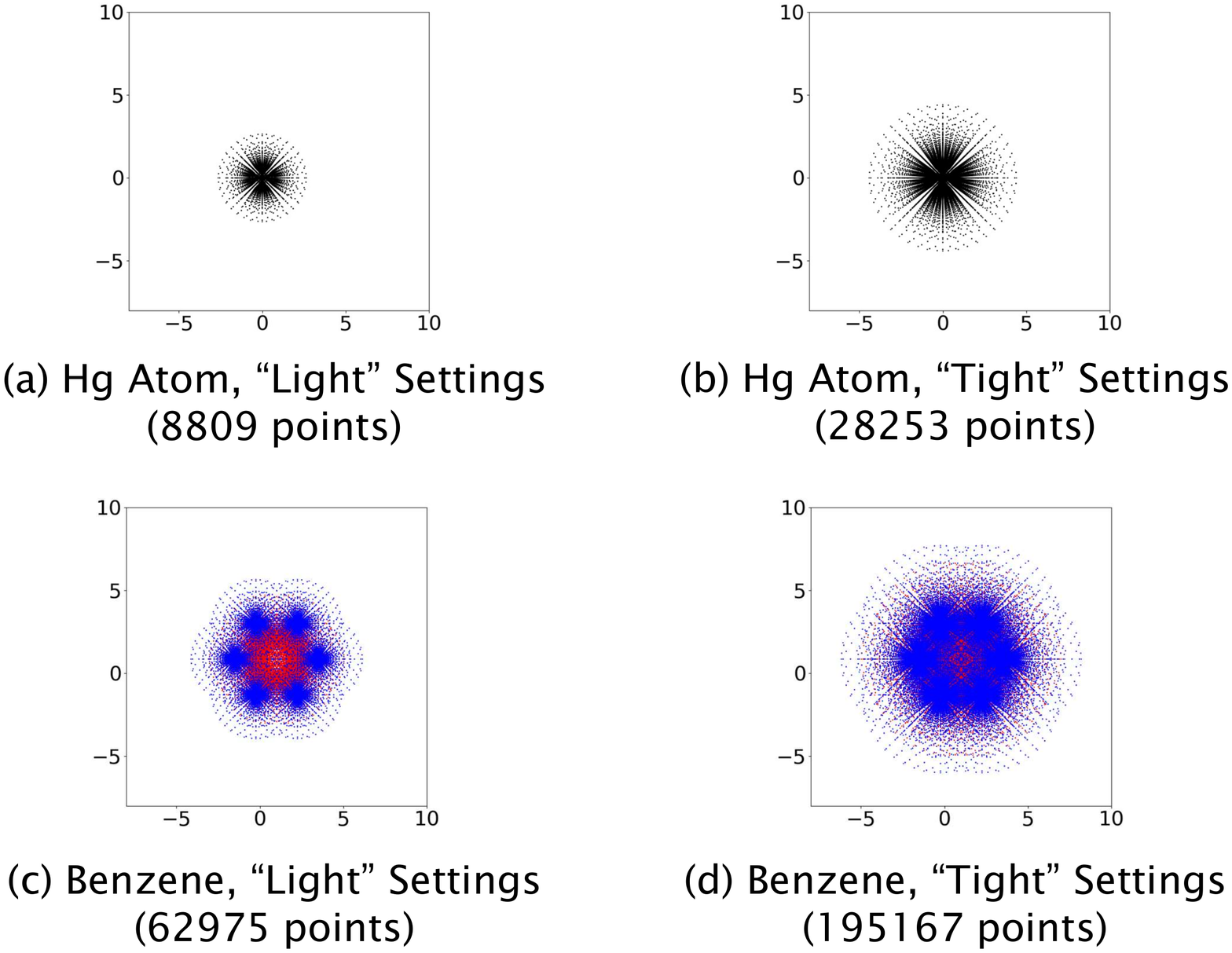}
  \caption{Visualization of real-space integration points for two simple molecules:  a mercury atom with (a) ``light'' and (b) ``tight'' integration settings and a benzene molecule with (c) ``light'' and (d) ``tight'' integration settings.  The number of integration points contained in each subfigure is listed in the caption.  In subfigures (c) and (d), integration points generated from the carbon and hydrogen atoms are marked red and blue, respectively.  Units are in bohr.}
  \label{fig:IntPoint}
\end{figure}

A visualization of the real-space integration grids in FHI-aims for a mercury atom is shown in Figures \ref{fig:IntPoint}a (light settings of FHI-aims) and \ref{fig:IntPoint}b (tight settings of FHI-aims.  Analogous figures for the multi-atom benzene molecule are shown in Figures \ref{fig:IntPoint}c and \ref{fig:IntPoint}d.  The grid consists of spherical shells of points around each nucleus, with individual grid points located on each shell according to the point distributions described by Lebedev \emph{et al}.~\cite{VILebedev75,VILebedev76,VILebedev99} and Delley~\cite{BDelley96}.  For the mercury atom, the outermost spherical shells can be observed as ``rays'' of points far from the respectively nuclei.  Closer to the nucleus, there are less points per spherical shell, but the radial density of spherical shells increases systematically (see Eq. 18 in~\cite{Blum09}) to account for rapidly varying wavefunctions.

In polyatomic systems like benzene, every atom contributes integration points to the overall integration grid of the system.  The radial cutoffs for the radial integration grids well exceed bond lengths in molecules and materials, leading to a densely-interlocking, irregular cloud of integration points covering both nuclear and interstitial regions.  Integration weights $w(\bm{r})$ are calculated on-the-fly (see Eq. (111) and Appendix C in~\cite{FKnuth15} for the exact definition of the respective weight functions used in FHI-aims), yielding a form for the real-space Hamiltonian integration
\begin{equation}
\begin{split}
  h_{ij} &= \sum_{\bm{r}} w(\bm{r})\varphi^{*}_{i}(\bm{r})\hat{h}_{KS}\varphi_{j}(\bm{r})
\end{split}
\end{equation}
or (for periodic systems)
\begin{equation}
\begin{split}
  h^{uc}_{ij} &= \sum_{\bm{r}} w(\bm{r})\varphi^{*}_{i}(\bm{r})\hat{h}_{KS}\varphi_{j}(\bm{r}) \, .
\label{eq:H_int_aims}
\end{split}
\end{equation}

FHI-aims calculates $w(\bm{r})$ using a partition-of-unity approach~\cite{ADBecke88} with a modified form of the $O(N_{atom})$ partitioning scheme proposed by Stratmann \emph{et al}.~\cite{REStratmann96} (see Eqs. 111-112 and C.5-C.8 in Knuth \emph{et al}.~\cite{FKnuth15}), where $N_{atom}$ is the number of atoms in the computational cell.  Here, it is sufficient to note that the resulting set of points is not an simple even-spaced grid and may be freely distributed across processes. Also, the process to update overlap matrix integrals $s_{ij}$ or $s_{ij}^{uc}$ (once per SCF cycle) is exactly analogous to the process outlined below for the elements of the Hamiltonian.

The default basis sets \{$\varphi_{i}(\bm{r})$\} used by FHI-aims for evaluating Eq. (\ref{eq:H_int_aims}) are preconstructed sets of numerical atom-centered orbitals (NAO)  optimized for KS-DFT-based total energy calculations, as outlined in~\cite{Blum09}  The NAO basis sets used in FHI-aims have been shown to have accuracy on par with the best available benchmark codes for total and atomization energies of molecules~\cite{SRJensen17} and calculated equations of state and band energies for solids~\cite{lejaeghere2016reproducibility,WPHuhn17}.  A cutoff potential is used to smoothly limit the spacial extent of the basis functions.  The cutoff potential (Eq. (9) in~\cite{Blum09}) reaches infinity at a user-definable outermost edge that is 6~{\AA} for most chemical species using tight settings.  Thus, any radial functions approach zero smoothly at this radius and remain zero at larger distances from their center.

\subsection{Real-Space Domain Decomposition (RSDD)}

\label{sec:batch_algo}

The RSDD is used for the update of the electron density and integration of various matrix elements for semi-local operators.  We will focus on integration of the real-space Hamiltonian matrix (Eqs. (\ref{eq:RealSpaceHamNonPeriodic}) and  (\ref{eq:RealSpaceHamPeriodic})) in this section, although the procedure presented applies to matrix elements of an arbitrary semi-local operator.  Additional steps related to the electron density update will be presented when relevant.  A detailed study of the expected linear scaling in runtime of the RSDD algorithm, including an examination of the performance of various partitioning schemes for integration points, was performed by Havu \emph{et al}~\cite{Havu09}.

\begin{figure}
  \centering
  \includegraphics[trim={7cm 1cm 8cm 1cm}, clip, width=0.8\textwidth]{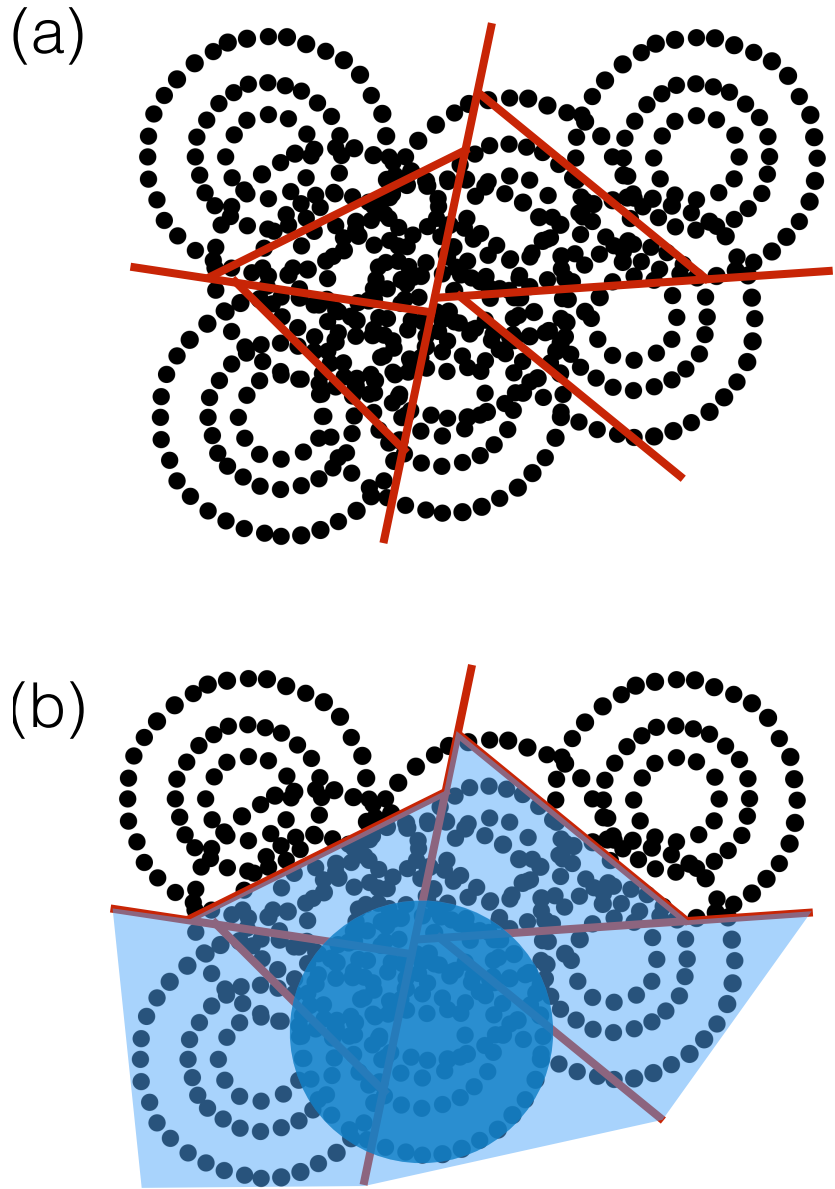}
    \caption{A visualization of the batch partitioning scheme used by FHI-aims.  Black points denote integration points, and red lines denote boundaries between batches of points.  In subfigure (b), the dark blue circle denotes the support of a local basis function, and batches on which the local basis function has non-zero support are highlighted in light blue.}
  \label{fig:Grid_Partitioning}
\end{figure}

\subsubsection{Grid Partitioning}

Figure \ref{fig:Grid_Partitioning}a shows a visualization of the partitioning of the set of integration points $P$ into mutually-disjoint ``batches'' of points $B_{\nu} \subset P$.  Each batch is a compact, spatially-local set of points on which vectorized operations will be performed and are distributed in a round-robin fashion among processes.  At no time does any individual process have knowledge of the full real-space integration grid.

The partitioning is accomplished by using a ``grid adapted cut-plane method'' (Algorithm 3 of~\cite{Havu09}) in which the full set $P$ is iteratively divided to smaller subsets using cut-planes (red lines in \ref{fig:Grid_Partitioning}a) until some targeted number of points per subset is reached.  The resulting subsets are the batches $B_{\nu}$.  The targeted number of points per batch will be denoted as the ``batch size'' throughout this paper.  The default batch size in FHI-aims is 100 points per batch for CPU-only calculations and 200 points per batch for GPGPU-accelerated calculations.  We increase the batch size for GPGPU-accelerated calculations as this will decrease the number of batches, decreasing the amount of CPU-GPU communication and increasing the amount of work accelerated per batch.  In our tests, increasing the batch size beyond 200 points does not lead to further overall acceleration.

\subsubsection{Parallelization and Dimensionality Reduction}

Having generated a set of mutually disjoint batches $\{B_{\nu}\}$, we may rewrite Eq. (\ref{eq:H_int_aims}) as
\begin{equation}
  h^{uc}_{ij} = \sum_{\nu}h^{uc}_{ij}[B_{\nu}] 
  \label{eq:H_decomp}
\end{equation}
where
\begin{equation}
  h^{uc}_{ij}[B_{\nu}] = \sum_{\bm{r}\in B_{\nu}} w(\bm{r})\varphi^{*}_{i}(\bm{r})\hat{h}_{KS}\varphi_{j}(\bm{r})
  \label{eq:H_contrib_batch}
\end{equation}
is the contribution of batch $B_{\nu}$ to the real-space Hamiltonian matrix element $h^{uc}_{ij}$.  Eqs. (\ref{eq:H_decomp}) and (\ref{eq:H_contrib_batch}) are trivially parallelizable over batches.  The RSDD algorithm parcels out batches to MPI tasks, so that each MPI task owns multiple batches, and each batch is owned uniquely by a single MPI task~\cite{Havu09}.

The indices in Eqs. (\ref{eq:H_decomp}) and (\ref{eq:H_contrib_batch}) by default run over all basis functions.  $h^{uc}_{ij}[B_{\nu}]$ for a given batch $B_{\nu}$ will be sparse for medium-to-large-sized systems, as many basis functions are centered too far away to touch the integration points in $B_{\nu}$.  While it may be tempting to store $h^{uc}_{ij}[B_{\nu}]$ in a sparse format, the usage of a sparse matrix format will introduce an intermediate non-trivial indexing step when accessing the matrix elements, impeding vectorization due to indirect addressing.  More importantly, the sparse representations for matrices $h^{uc}_{ij}$ and $h^{uc}_{ij}[B_{\nu}]$ will scale in size as $O(N_{atom})$.  Retaining local copies of the full $h^{uc}_{ij}[B_{\nu}]$ on each separate MPI process will eventually become the dominant memory cost of the calculation.

\begin{figure}
  \centering
  \includegraphics[trim={2cm 3cm 2cm 2cm},clip,width=0.8\textwidth]{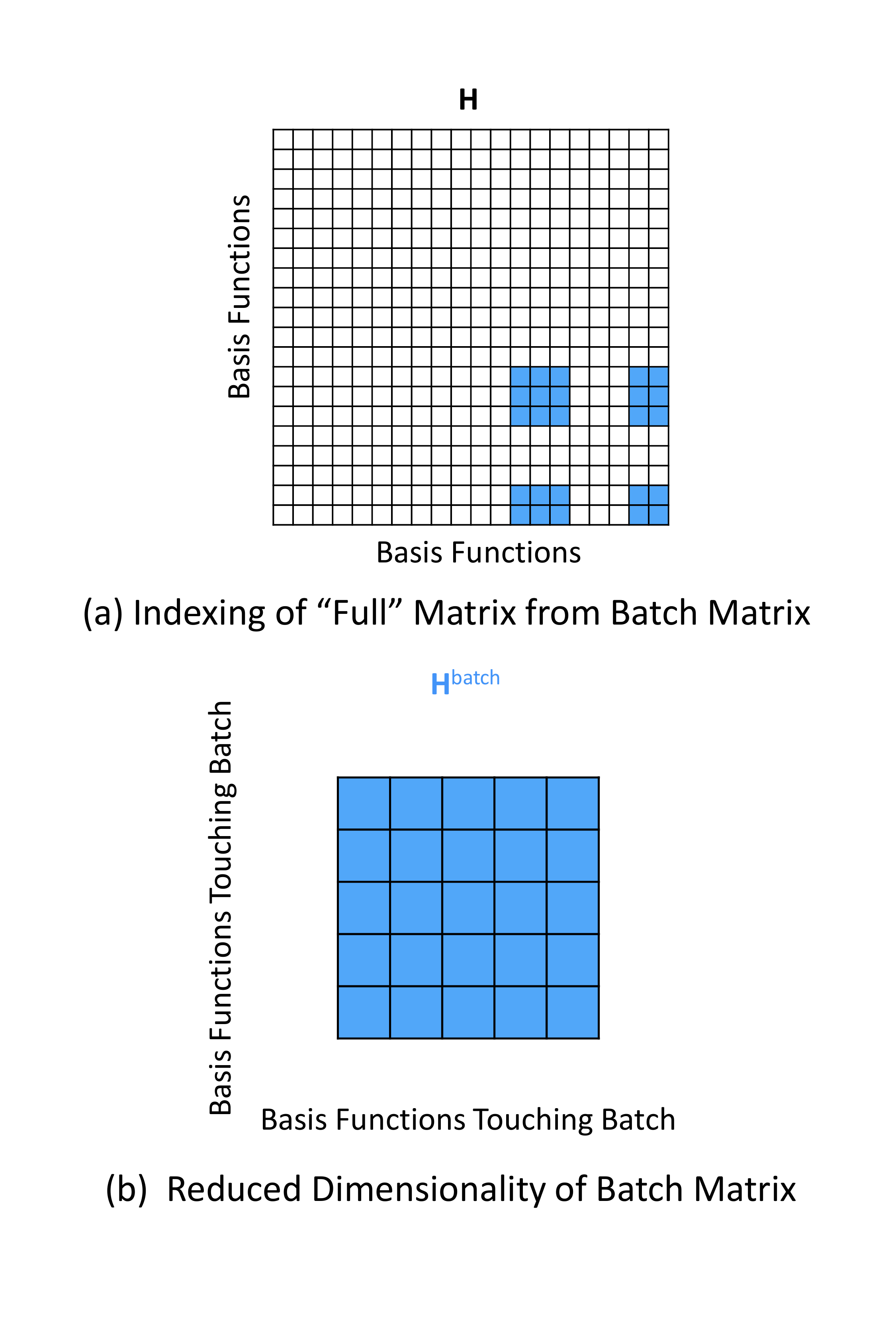}
    \caption{The contribution of a batch to (a) the real-space Hamiltonian matrix and (b) the batch Hamiltonian matrix.   Blue squares indicate matrix elements $h^{uc}_{ij}$ to which the batch may contribute, and white squares denote matrix elements with no contribution from the batch.  We show the real-space Hamiltonian matrix as an uncompressed full matrix for simplicity; see Figure \ref{fig:Matrix_Layout_Visualization} for the matrix layouts used in FHI-aims.}
  \label{fig:Batch_Matrix_Visualization}
\end{figure}

For each batch, we instead define a reduced-dimensional subspace $\mathrm{nnz}(B_{\nu})$ consisting of basis functions with support on the batch, i.e., basis functions that are nonzero on at least part of the grid points in this batch.  $h^{uc}_{ij}[B_{\nu}]$ can be considerably reduced in size by calculating only matrix elements between basis elements $\varphi_{i}$ lying in $\mathrm{nnz}(B_{\nu})$, yielding a dense ``batch Hamiltonian matrix'' with form identical to Eq. (\ref{eq:H_contrib_batch}).  We take $h^{uc}_{ij}[B_{\nu}]$ to refer to this subspace batch Hamiltonian matrix for the rest of this paper.  When evaluating the electron density update (Eq. (\ref{eq:Density})) using the RSDD algorithm, an analogous subspace representation is employed for the density matrix $n_{ij}[B_{\nu}]$ and (if required) for the energy-weighted density matrix $q_{ij}[B_\nu]$ when calculating Pulay forces (Eq. (\ref{eq:forces})) and the stress tensor (Eq. (\ref{eq:stress_tensor})).

Defining the $\|\mathrm{nnz}(B_{\nu})\|~\times~\|B_{\nu}\|$ matrix,
\begin{equation}
  K_{i\bm{r}} \equiv w(\bm{r})\varphi_{i}^{*}(\bm{r}),
\end{equation}
and the $\|B_{\nu}\|~\times~\|\mathrm{nnz}(B_{\nu})\|$ matrix
\begin{equation}
  L_{\bm{r}j} \equiv  \hat{h}_{KS}\varphi_{j}(\bm{r}),
\end{equation}
all matrix elements of $h^{uc}_{ij}[B_{\nu}]$ may be evaluated simultaneously by rewriting Equation \ref{eq:H_contrib_batch} as a matrix multiplication
\begin{equation}
  h^{uc}_{ij}[B_{\nu}] = \sum_{\bm{r}\in B_{\nu}}K_{i\bm{r}}L_{\bm{r}j}.
  \label{eq:batch_matrix_mult}
\end{equation}

A visualization of the difference between $h^{uc}_{ij}$ and $h^{uc}_{ij}[B_{\nu}]$ is provided in Figure \ref{fig:Batch_Matrix_Visualization}.  $h^{uc}_{ij}$ is a sparse matrix for large systems and localized basis sets, since only a bounded number of basis functions in some proximity to one another will overlap.  $h^{uc}_{ij}[B_{\nu}]$, in contrast, is dense for sufficiently compact set of points, e.g., a sufficiently small batch, since its indices only run over the subset of basis functions that are non-zero anywhere within the compact set of grid points (i.e., these basis functions are already located close to one another). It therefore has a memory consumption of $\|\mathrm{nnz}(B_{\nu})\|^{2}$.

While different batches will have different values for $\|\mathrm{nnz}(B_{\nu})\|$ depending on local basis set density and geometry, there exists some upper limit $\mathrm{max}(\|\mathrm{nnz}(B_{\nu})\|)$ for the number of basis functions interacting with any batch.  $\mathrm{max}(\|\mathrm{nnz}(B_{\nu})\|)$ depends on basis set density and local geometry \emph{but not on the system size}.  The independence of the size of each batch from the overall system size is responsible for the observed $O(N_{atom})$ execution of integrals, as shown in Havu \emph{et al.}~\cite{Havu09}.

Once $h^{uc}_{ij}[B_{\nu}]$ has been calculated, the local copy of $h^{uc}_{ij}$ on each MPI task (here denoted $h^{uc,task}_{ij}$) is updated.  There are two matrix formats for the local copy of $h^{uc}_{ij}$ commonly used in FHI-aims: a globally-indexed compressed sparse row (CSR) format and a locally-indexed dense format.Both are further explained in two subsections below. The term ``globally indexed'' refers to storage of an identical, sparse copy of the final matrix $h^{uc}_{ij}$ across all MPI tasks. In the ``globally indexed'' case, the size of the stored array thus grows with system size on each MPI task, regardless of the number of MPI ranks employed. In contrast, the term ``locally indexed'' refers to the storage on each MPI task of separate, dense matrix versions of only those integration contributions to $h^{uc}_{ij}$ that are non-zero on the localized subset of grid points handled by that task. In the ``locally indexed'' case, the size of the stored array on each task thus remains bounded if the system size increases, as long as the number of MPI ranks is increased along with the system size.

\begin{figure}
  \centering
  \includegraphics[trim={0cm 8cm 0cm 5cm},clip,width=0.8\textwidth]{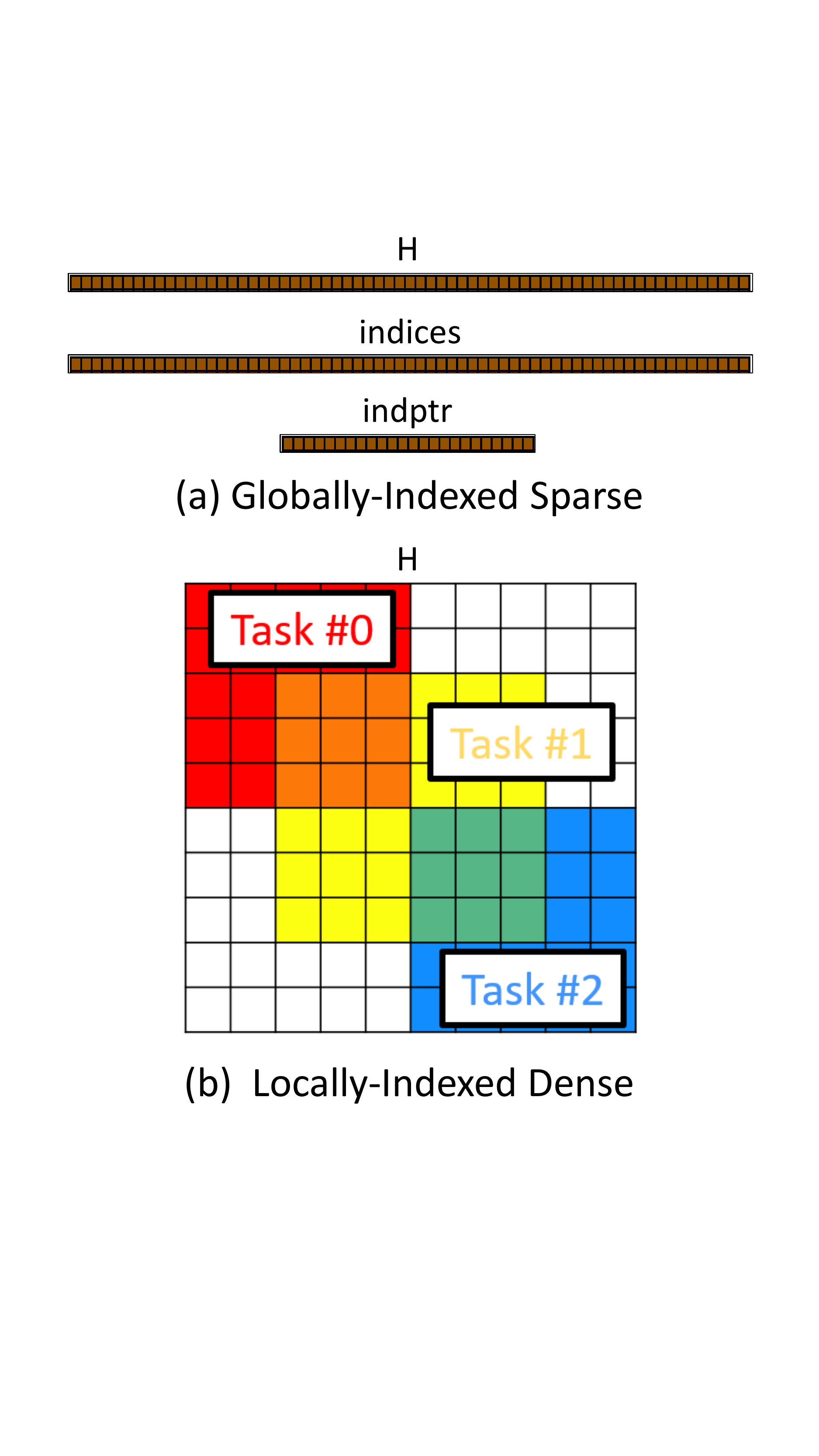}
    \caption{Visualization of the two main matrix formats used for storing real-space matrices in FHI-aims.  Subfigure (a) shows the globally-indexed sparse format, where each MPI task has a full copy of the matrix H in a sparse format (here, CSR) along with two indexing arrays indices and indptr.  Subfigure (b) shows the locally-indexed dense format, where each MPI task has a partial section of the matrix in a dense format.}
  \label{fig:Matrix_Layout_Visualization}
\end{figure}

\subsubsection{Globally-Indexed Sparse Real-Space Hamiltonian}

Shown in Figure \ref{fig:Matrix_Layout_Visualization}a is a visualization of the globally-indexed sparse matrix format, in which each MPI task has a full copy of the real-space matrix along with associated indexing arrays.  In FHI-aims, we use the CSR format for the sparse representation.  Indexing from $h^{uc}_{ij}[B_{\nu}]$ to $h^{uc}_{ij}$ in the globally-indexed CSR format requires two steps.  First, the basis function index in $\mathrm{nnz}(B_{\nu})$ is mapped back to the basis function index in the set of all basis functions, i.e. the row/column indices of the matrix in Figure \ref{fig:Batch_Matrix_Visualization}b are mapped back to the row/column indices of the matrix in Figure \ref{fig:Batch_Matrix_Visualization}a.  Second, the matrix element is mapped back to the sparse matrix representation used to store the real-space Hamiltonian matrix via indexing arrays.

Each process, after evaluating all batches assigned to it, has a local incomplete copy of the real-space Hamiltonian matrix 
\begin{equation}
  h^{uc,\mathrm{task}}_{ij} = \sum_{B_{\nu}~\mathrm{on~task}}h^{uc}_{ij}[B_{\nu}].
  \label{eq:H_task}
\end{equation}
At the end of the RSDD algorithm, an in-place synchronization call sums up all local copies of $h^{uc}_{ij}$ across processes,
\begin{equation}
  h^{uc}_{ij} = \sum_{\mathrm{tasks}}h^{uc,\mathrm{task}}_{ij},
  \label{eq:H_task_2}
\end{equation}
yielding the final values for the real-space Hamiltonian matrix synchronized across all processes.  This is the only communication step in the RSDD algorithm.

The sparse matrix storage used in this matrix format makes it an efficient choice for smaller calculations.  However, it has two key disadvantages impeding its usage for large-scale calculations and massively-parallel architectures.  The sparse storage of $h^{uc}_{ij}$ nevertheless scales as $O(N_{atom})$ on each MPI task and will be a memory bottleneck for large-scale systems, in particular ill-suited to reside in the on-board GPGPU memory.  Additionally, the usage of non-trivial indexing arrays considerably impedes vectorization, making it ill-suited to the massively-vectorized GPGPU programming paradigm.

\subsubsection{Locally-Indexed Dense Real-Space Hamiltonian}

Shown in Figure \ref{fig:Matrix_Layout_Visualization}b is a visualization of the locally-indexed dense matrix format, where $h^{uc}_{ij}$ is distributed across processes.  Each MPI task stores its portion of the integrals that contribute to $h^{uc}_{ij}$ locally in a dense format -- i.e., those partial integrals $h^{uc,\mathrm{task}}_{ij}$ (Eq.~\ref{eq:H_task}) that are non-zero for this task. In short, only matrix elements involving basis functions with non-zero support on at least one batch on a given process, i.e. with support on the set $\cup_{B_{\nu}~on~task}B_{\nu}$, are stored.  The indexing step from $h^{uc}_{ij}[B_{\nu}]$ to $h^{uc,\mathrm{task}}_{ij}$ is amenable to GPGPU resources due to the dense storage of $h^{uc,\mathrm{task}}_{ij}$, amounting to a usage of a simple pre-calculated look-up table.  

After the RSDD algorithm has concluded, the locally-index matrix elements $h^{uc,\mathrm{task}}_{ij}$ are distributed across MPI tasks in an overlapping fashion, i.e. a given matrix element $h^{uc}_{ij}$ may have non-zero contributions $h^{uc,\mathrm{task}}_{ij}$ on multiple MPI tasks.  Implementing Eq. (\ref{eq:H_task_2}) to convert these local matrices into a form $h_{ij}$ (non-periodic systems) or $h_{ij}(\bm{k})$ (periodic systems) suitable for the eigensolver (here, BLACS for ELPA) is considerably more tedious than in the globally-indexed case, as book keeping to determine ownership of matrix elements and point-to-point communication between MPI tasks is required.  Nevertheless, this synchronization step only occurs once and has negligible effect on the overall timings.

This dense matrix format consumes more memory than the globally-indexed sparse format for small calculations, as negligible matrix elements will be computed and stored in the local dense matrices.  For sufficiently large calculations with sufficiently many MPI tasks, the distributed nature of this format is far more efficient in overall memory usage than the globally-indexed sparse format, and the resulting local dense matrices fit completely into GPGPU onboard memory, minimizing CPU-GPGPU communication.

\subsubsection{Density and Density Derivatives, Force and Stress Tensor Components}

In principle, all computational steps of other quantities derived from basis functions on the real-space grid can profit from very similar sparsity considerations as laid out for the Hamiltonian and overlap matrices above. This entails the density and its derivatives, which (cf. Eq.~(\ref{eq:Density})) obey precisely the same locality constraints in each batch of grid points as the Hamiltonian and overlap matrices. The force and stress tensor components are likewise cast in the form of numerical integrals that, within each batch, are touched by only a bounded number of localized basis functions in the limit of large systems. 

The computationally dominant operations for all these terms are, again, dense matrix multiplications. Since they can be carried out on the same set of distributed batches of grid points as the Hamiltonian and overlap matrices, the relevant index ranges (grid points in each batch and non-zero basis functions in each batch) are effectively identical to those in Eq.~(\ref{eq:batch_matrix_mult}). Quantities are either kept on the distributed grid directly (density and its derivatives, which are local quantities) or (for forces and stress tensor components) the necessary integrals can be carried out by initially assembling matrix elements with basis functions as their indices. The process for the latter matrix elements is essentially identical to the detailed distribution strategy described above for the Hamiltonian matrix. Actual force and stress tensor components in terms of atomic coordinates can then be summed up from these matrix elements once, after the entire integration grid has been completely processed.

\subsection{GPGPU Acceleration of the Real-Space Domain Decomposition Algorithm}
\label{sec:GPGPU_Accel}

\begin{figure*}
  \centering
  \includegraphics[trim={5cm 3.5cm 6cm 0cm}, clip, width=\textwidth]{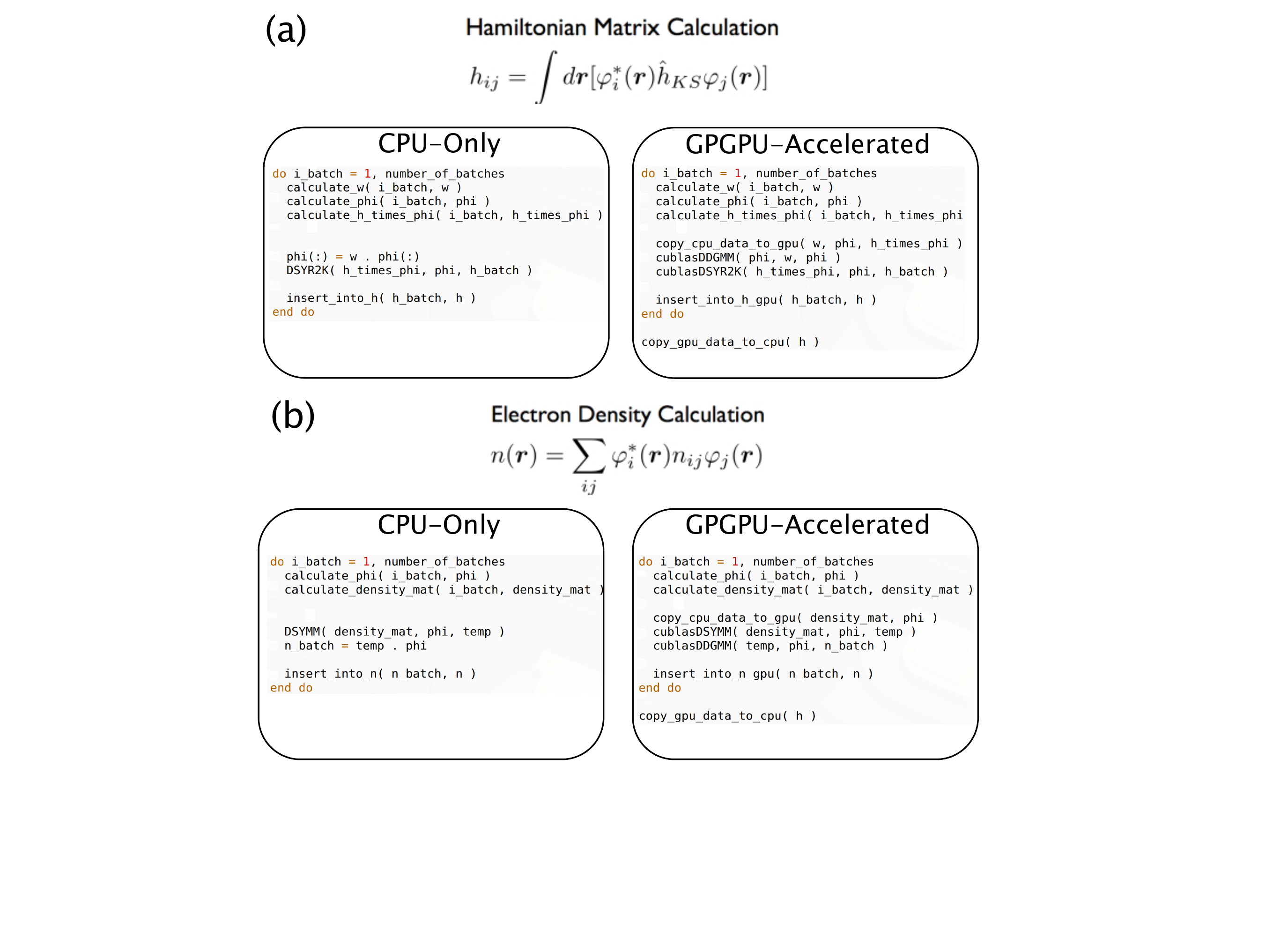}
    \caption{Pseudocode comparing CPU-only and GPGPU-accelerated implementations of the RSDD algorithms in FHI-aims for the (a) real-space Hamiltonian matrix calculation and (b) electron density calculation.  The evaluation of the tau matrix for meta-GGAs has been omitted in (a) for brevity.  It is calculated via a single DGEMM/cublasDDGMM call and added to the $h^{uc}_{ij}$ matrix at the end of each loop iteration. In (a), the final summation of all partial contributions to $h^{uc}_{ij}$ across different MPI tasks happens either immediately after the pseudocode shown (globally indexed version) or later (locally indexed version), when the full matrix elements $h_{ij}(\bm{k})$ of Eq. (\ref{eq:HamPeriodic}) are constructed from Bloch phase factors and partially summed versions of $h^{uc}_{ij}$ stored on each separate MPI task.}
  \label{fig:Pseudocode}
\end{figure*}

Figure \ref{fig:Pseudocode} shows pseudocode for the CPU-only and GPGPU-accelerated implementations of the real-space (a) Hamiltonian matrix integration and (b) electron density update.  We refer to these two operations collectively as ``the real-space operations'' hereafter.  The basic structure of the real-space operations consists of a loop over every batch assigned to the process, possibly followed by a final post-processed synchronization across all processes for integrals.  No synchronization is necessary for the electron density update, as the real-space points where the electron density is defined are distributed across processes.  For each batch assigned to a process, its contribution to the desired quantity is calculated.  The processing of each batch can be broken up into three phases:  initial processing where the necessary quantities (basis functions on each grid point, their gradients, etc.) are computed, a sequence of serial dense linear algebra operations (e.g., Eq. (\ref{eq:batch_matrix_mult})), and a final indexing step to sort matrix contributions from each batch into their respective storage arrays on each MPI task.

Common to the initial processing phase for all real-space operations is the construction of the basis elements $\varphi_{i} \in \mathrm{nnz}(B_{\nu})$ for all $\bm{r} \in B_{\nu}$.  The initial processing phase for the electron density update additionally includes the reduction of the density matrix of the full system (stored in a sparse representation) to the dense-but-smaller $n_{ij}$ matrix as outlined in the previous section.  The real-space Hamiltonian matrix integration requires that the pre-computed integration point weights be re-indexed for the current batch and the quantities $\{\hat{h}_{KS}\varphi_i(\bm{r})\}$ be evaluated on every point in $B_{\nu}$ for all basis elements in $\mathrm{nnz}(B_{\nu})$.  

The construction of $\{\varphi_i(\bm{r})\}$, and $\{\hat{h}_{KS}\varphi_i(\bm{r})\}$ for each point on a batch is executed per point, with relatively small workload compared to the final DGEMMs, but with some thread-divergent branching statements during the construction, i.e. not immediately suitable for GPGPUS.  As we will see below, for the integrals, we can overlay this work on the CPU while the GPU works on something else.

The dense linear algebra phase consists of two subroutine calls for Hamiltonian integration: a matrix multiplication (DSYR2K for the real-space Hamiltonian matrix and DSYMM for the electron density) and a dot product.  This step is offloadable to the GPGPU at the cost of communication of the quantities previously calculated during the initial processing phase:  the integration weights $w(\bm{r})$, basis functions $\{\varphi_i(\bm{r})\}$, and $\{\hat{h}_{KS}\varphi_i(\bm{r})\}$ matrix for the real-space Hamiltonian integration in Eq. (\ref{eq:H_int_aims}), and the basis functions $\varphi$ and batch subspace restricted density matrix $n_{ij}$ for the electron density in Eq. (\ref{eq:Density}).

The final phase for each loop iteration is an indexing phase, where the results from the dense linear algebra phase $h^{uc}_{ij}[B_{\nu}]$ are indexed back into the accumulated matrices $h^{uc}_{ij}$.  The GPGPU acceleration strategy for the real-space Hamiltonian integration diverges based on the choice of indexing for the Hamiltonian matrix $h^{uc}_{ij}$:  

(i) The globally-indexed CSR format uses several thread-divergent branching conditionals, complicating GPGPU acceleration of the indexing.  The GPGPU-accelerated algorithm thus communicates the results of the dense linear algebra step for each batch back to the CPU, which performs the indexing.  In this approach, the CPU is idle while the GPGPU evaluates the dense linear algebra for each batch.

(ii) Alternatively, when using the locally-indexed dense storage format for $h^{uc}_{ij}$, indexing is embarrassingly parallelizable, allowing for the usage of custom CUDA kernels to perform the indexing on the GPGPU.  After the CPU has communicated $\{\varphi_i(\bm{r})\}$ and $\{\hat{h}_{KS}\varphi_i(\bm{r})\}$ to the GPGPU, it is free to begin calculating $\{\varphi_i(\bm{r})\}$  and $\{\hat{h}_{KS}\varphi_i(\bm{r})\}$ for the next batch, allowing for effective overlay of computation between the CPU and GPGPU.  Communication from the GPGPU back to the CPU occurs only once at the end of the algorithm, when the GPGPU communicates its copy of $h^{uc}_{ij}$ back to the CPU.

A similar approach to (ii) is employed when calculating the electron density or its derivatives, as the mapping from real-space points $\bm{r} \in B_{\nu}$ to the set of all integration points assigned to the present process $\cup_{B_{\nu}~on~task}B_{\nu}$ is trivial.  This permits the electron density indexing to be performed via a simple CUDA kernel and stored on the GPGPU locally, similarly avoiding communication from the GPGPU back to the CPU communication for each batch.  The electron density stored on the GPGPU is communicated back to the CPU at the end of the algorithm.

\section{Results}

All benchmark calculations were performed with the full-potential, all-electron FHI-aims electronic structure code~\cite{Blum09,Havu09} using its production-quality ``tight'' basis sets and ``tight'' real-space integration grids and Hartree potential.  The PBE functional~\cite{PerdewBurkeErnzerhof96} was used to model exchange-correlation effects.  We use a $\Gamma$-point-only k-grid in this work, as we focus on real-space operations whose timings are independent of the selected reciprocal-space grid.  All timings shown use the locally-indexed dense matrix format (Figure \ref{fig:Matrix_Layout_Visualization}b) for the Hamiltonian integration and forces/stress evaluation. The choice of ``tight'' settings is important since the larger workload associated with denser grids and larger basis sets can dominate the overall effort associated with a particular DFT-based simulation project. By construction, larger workloads of this kind also benefit more from GPGPU acceleration. For progressively lighter settings, smaller but still useful speedups would be expected. The benchmarks below reflect the highest-cost parts of a routine simulation where, it turns out, GPU acceleration will most efficiently take off the edge of a very large workload.

The time-intensive operations for semi-local DFT in FHI-aims are the real-space Hamiltonian integration (Eq. (\ref{eq:RealSpaceHamNonPeriodic}) or (\ref{eq:RealSpaceHamPeriodic})), density update (Eq. (\ref{eq:Density})), Hartree potential calculation (Eq. (\ref{eq:HartreePotential})), and the solution of the Kohn-Sham eigenvalue equation (Eq. (\ref{eq:KS-EV-Prob})).  

Figure \ref{fig:SCF_Cycle} shows a schematic for the program flow of a standard KS-DFT calculation with geometry relaxation or molecular dynamics.  First, the electronic structure for current geometry is converged by iterative calculation of the Hartree potential (\ref{eq:HartreePotential}), Hamiltonian operator (Eq. (\ref{eq:Hamiltonian})), Hamiltonian matrix elements (Eq. (\ref{eq:RealSpaceHamNonPeriodic}) or \ref{eq:HamPeriodic}), and electron density via the Kohn-Sham eigenvalue equation (Eqs. (\ref{eq:Density}) and (\ref{eq:KS-EV-Prob})).  We denote a single collective iteration of these operations in which the electronic structure is updated as a ``standard'' SCF iteration.

Once self-consistency is reached in the electronic structure, $\forceat$ and $\stresstens$ may be calculated by Eqs. (\ref{eq:forces}) and (\ref{eq:stress_tensor}) respectively.  The implementation of Eqs. (\ref{eq:forces}) and (\ref{eq:stress_tensor}) in FHI-aims are published in ~\cite{Blum09,FKnuth15}; here, it is sufficient to highlight two contributions to $\forceat$ and $\stresstens$ which contribute appreciably to timings:
\begin{itemize}
\item The Hellmann-Feynman contribution to $\forceat$ and $\stresstens$ is calculated alongside the Hartree multipole summation, as it relies on the Hartree potential of the system.  Thus, the domain decomposition strategy is not used for evaluating this contribution, and GPGPU acceleration is not employed.
\item The Pulay contribution to $\forceat$ and $\stresstens$ is calculated alongside the density update, as it relies on the density matrix $n_{ij}$ and the energy-weighted density matrix $q_{ij}$.  The domain decomposition strategy is used for evaluating this contribution.  Its evaluation is GPGPU-accelerated using the strategy outlined in Section \ref{sec:GPGPU_Accel}.
\end{itemize}
While the calculations of these quantities are computationally expensive, they are ideally evaluated only once per geometry step (outer loop of Figure \ref{fig:SCF_Cycle}), whereas the SCF iterations will be evaluated multiple times per geometry step (inner loop of Figure \ref{fig:SCF_Cycle}).  

In principle, the electronic structure should be further iterated until $\forceat$ and $\stresstens$ have converged.  The production settings for FHI-aims are sufficiently tight that in practice only a single calculation of $\forceat$ and $\stresstens$ are needed at the end of each self-consistency cycle.  The geometry is then updated using $\forceat$ and $\stresstens$.

Accordingly, we benchmark three different types of SCF iterations in this paper:
\begin{itemize}
    \item ``SCF'': a single iteration of the SCF cycle loop,  
    \item ``SCF + $\forceat$'': a single iteration of the SCF cycle loop followed by the computation of forces $\forceat$, and 
    \item ``SCF + $\forceat$ + $\stresstens$'': a single iteration of the SCF cycle loop followed by the computation of forces $\forceat$ and stress tensor $\stresstens$.
\end{itemize}
We emphasize that these timings are per iteration, not per cycle.  As outlined in Figure \ref{fig:SCF_Cycle}, a given system geometry loop will contain multiple iterations of ``SCF'' but ideally only one iteration of ``SCF + $\forceat$'' or ``SCF + $\forceat$ + $\stresstens$'' per geometry step.

We cover a broad range of GPGPU-accelerated architectures by performing benchmarks on four different architectures provided by three different computing clusters:  a local cluster (``timewarp'') at Duke University, a dedicated testing cluster (``PSG'') at NVIDIA Corporation, and the Lassen supercomputer at Lawrence Livermore National Laboratory (``LLNL'').  The hardware architectures, which we categorize based on CPU and GPU generations, are IvyBridge/GP100 [timewarp], Haswell/P100 [PSG], Skylake/V100 [PSG], and POWER9/V100 [LLNL].  More information on architectures may be found in Table \ref{tab:archs}.  Full node utilization was employed for all architectures except on IvyBridge/GP100 [timewarp], where 16 MPI tasks were used (out of the 20 physical cores) due to limitations imposed on Pascal GPGPUs by the NVIDIA MPS service, which was used to distribute the computational load to the GPGPU.

\begin{table*}
\centering
\scriptsize
\begin{tabular}{|c|c|c|c|c|}
\hline
Node Type           & IvyBridge/GP100 & Haswell/P100 & Skylake/V100 & POWER9/V100\\ 
\hline
\hline
Computer Cluster    & timewarp & PSG  & PSG & LLNL's Lassen\\
\hline
CPU                 & 2x Intel Xeon & 2x Intel Xeon  & 2x Intel Xeon & 2x IBM POWER9 \\
                    & E5-2670v2     & E5-2698v3      & Gold 6148     & AC922 \\
                    & (20 cores,    & (32 cores,     & (20 cores,    & (40 cores, \\
                    & Ivy Bridge)   & Haswell)       & Skylake)      & POWER9) \\
\hline
GPGPU               & 1x Quadro GP100 & 4x Tesla P100 & 4x Tesla V100 & 4x Tesla V100 \\
                    & (Pascal)        & (Pascal)      & (Volta)       & (Volta) \\
\hline
MPI Tasks/GPGPUs    & 16/1 & 32/4 & 20/4 & 40/4 \\
\hline
Compilers/Libraries & ifort 14.0,     & ifort 17.0,   & ifort 17.0,   & XL 2019.02.07, \\
                    & MKL 11.1.1,     & MKL 11.3.3,   & MKL 11.3.3,   & ESSL 6.1, \\
                    & IMPI 4.1.3,     & IMPI 5.0.3,   & IMPI 5.0.3,   & Spectrum MPI, \\
                    & CUDA 8.0        & CUDA 9.1      & CUDA 9.1      & CUDA 9.2 \\
\hline
\end{tabular}
\caption{Per-node configurations for architectures used in this work.}
\label{tab:archs}
\end{table*}

We consider two types of timings for a given operation in this work, the timing $t_{CPU}$ for an operation when employing all available CPU cores on a given node using the Message Passing Interface (MPI), and the timings $t_{GPGPU}$ for an operation when GPGPU acceleration has been employed alongside full CPU node utilization.  The GPGPU-accelerated speedup, or simply ``speedup'', is then defined as
\begin{equation}
  s = t_{CPU}/t_{GPGPU}.
\end{equation}
Using this definition, identical timings for CPU-only and GPGPU-accelerated operations correspond to a speedup of s = 1.0.  We note specifically that this is a hard comparison, in that the GPU+CPU timings must be better than a computation using all CPU cores on a given, state-of-the-art node to show a net speedup.

\subsection{Materials Used}

Two different sets of materials are considered in this work:
\begin{itemize}
\item The first set of materials is a benchmark set comprising 103 different inorganic compounds (bulk solids), proposed in~\cite{WPHuhn17}  The compounds in this benchmark set span 66 chemical elements in 10 structural prototypes, providing a broad coverage of the structural and chemical diversity found in real-world condensed-matter simulations. 3$\times$3$\times$3 supercells of the primitive cell for each material were used, leading to cell sizes of 27, 54, and 108 atoms.  In addition to discussing averaged / aggregated data, we highlight timings for the 54-atom diamond Si supercell calculation as a particular example in the body of this text.  The benchmark set is investigated on a single node of three out of the four architectures listed in Table \ref{tab:archs}, namely the IvyBridge/GP100 [timewarp], Haswell/P100 [PSG], and Skylake/V100 [PSG] architectures.

\item The second set of materials consists of a 5$\times$5 supercell of a 2D Bi\subscript{2}Se\subscript{3} bilayer.  A vacuum thickness of 40~$\text{\AA}$ was used.  In a standard FHI-aims calculation, the vacuum thickness contributes to timings only through the reciprocal-space contribution of the Ewald summation~\cite{BDelley96_es,Blum09} used to evaluate the electrostatic potential to the Hartree potential.  Calculations for the \bise bilayer were performed on the POWER9/V100 [LLNL] architecture using 2 to 128 nodes.
\end{itemize}

\subsection{GPGPU Acceleration Across Materials and Architectures}

Table \ref{tab:Si_ops_skl_v100} shows a comparison of timings computed on the Skylake/V100 [PSG] architecture for diamond Si.  The evaluation of the electron density  and its gradients is the dominant contribution to the total time for all SCF iteration types.  The relative weight of the density calculation increases for SCF iterations involving $\forceat$ and $\stresstens$ as the evaluation of the Pulay contribution must also be performed in this step.  For an SCF + $\forceat$ + $\stresstens$ iteration, 95\% of the calculation time is spent calculating the density and the Pulay contribution to $\forceat$ + $\stresstens$.  The timing for Hartree summation weakly increases for SCF cycles involving $\forceat$ as the Hellmann-Feynman forces must be calculated, but this is a comparatively small expense relative to the time required to compute the Pulay terms.  In contrast, the Hamiltonian integration does not include $\forceat$ or $\stresstens$ components, leading to consistent timings across iteration types.

\begin{table*}
\centering
\scriptsize
\begin{tabular}{|c|c|c|c|c|}
\hline
Operation         & Timing & ~SCF~ & SCF          & SCF             \\ 
                  &  Type  &       & + $\forceat$ & + $\forceat$    \\
                  &        &       &              & + $\stresstens$ \\
\hline
\hline
Integration (Eq. (\ref{eq:RealSpaceHamPeriodic})) & $t_{CPU}$ (s)   & 10.4 &  10.3 &  10.4 \\
                  & $t_{GPGPU}$ (s) &  1.6 &   1.6 &   1.6 \\
\hline
Hartree Summation (Eq. (\ref{eq:HartreePotential})) & $t_{CPU}$ (s)   &  4.3 &  16.6 &  20.2 \\
+ Hellmann-Feynman (Eqs. (\ref{eq:forces}), (\ref{eq:stress_tensor})) &                 &      &       &       \\
\hline
Electron Density (Eq. (\ref{eq:Density}))& $t_{CPU}$ (s)   & 20.7 & 243.7 & 565.2 \\
+ Pulay (Eqs. (\ref{eq:forces}), (\ref{eq:stress_tensor})) & $t_{GPGPU}$ (s) &  5.3 &  29.2 &  54.5 \\
\hline
Total Time for Iteration & $t_{CPU}$ (s)   & 35.5 & 270.8 & 596.0 \\
                         & $t_{GPGPU}$ (s) & 11.6 &  47.7 &  76.5 \\
\hline
\end{tabular}
\caption{Timings for various CPU-only and GPGPU-accelerated operations computed on the Skylake/V100 [PSG] architecture for a 54-atom Si supercell (diamond structure).  Only $t_{CPU}$ is presented for the Hartree summation and Hellmann-Feynman contributions to $\forceat$ and $\stresstens$, as GPGPU acceleration is not employed for these calculations.  All timings were taken from a single calculation.}
\label{tab:Si_ops_skl_v100}
\end{table*}

For all three iteration types, when GPGPU acceleration is enabled, there is a consistent s=6.4 speedup in the Hamiltonian integration on the Skylake/V100 [PSG] architecture.  Differences in the GPGPU performance are observed when calculating the density, its gradients,and potentially the Pulay contributions to forces and stress.  Speedups of 3.9, 8.4, and 10.4 are observed in the density calculation on the Skylake/V100 [PSG] architecture for the standard SCF, SCF + $\forceat$, and SCF + $\forceat$ + $\stresstens$ iterations, respectively.  The differences in speedups arise from the increased workload of dense linear algebra performed when calculating the Pulay contribution to $\forceat$ and $\stresstens$.  The computationally more demanding iterations with more dense linear algebra exhibit greater GPGPU-accelerated improvements for timings in both relative and absolute terms.  Speedups for the total times of an iteration show a similar trend to the density update, with speedups of 3.1, 5.7, and 7.8 observed on the Skylake/V100 [PSG] architecture for the three SCF iteration types.  The density update dominates the runtimes for the iterations.

Table \ref{tab:Si_iter_archs} extends the analysis to the IvyBridge/GP100 [timewarp] and Haswell/P100 [PSG] architectures with a comparison of total times for the three types of iterations.  The three architectures show similar trends; speedups increase as the linear algebra workload increases, with speedups of 2.9, 4.4, and 5.2 observed for IvyBridge/GP100 [timewarp] and 2.5, 5.8, and 6.7 observed for Haswell/P100 [PSG].  The trends in speedups observed on Skylake/V100 [PSG] hold for other GPGPU-accelerated architectures.

\begin{table}
\centering
\scriptsize
\begin{tabular}{|c|c|c|c|c|}
\hline
Architecture    & Timing & ~SCF~ & SCF          & SCF             \\ 
                & Type   &       & + $\forceat$ & + $\forceat$    \\
                &        &       &              & + $\stresstens$ \\
\hline
\hline
IvyBridge/GP100 & $t_{CPU}$ (s)   & 77.7 & 545.0 & 1130.5 \\
{[timewarp]}    & $t_{GPGPU}$ (s) & 26.5 & 123.3 &  218.8 \\
\hline
Haswell/P100    & $t_{CPU}$ (s)   & 38.5 & 315.2 &  682.0 \\
{[PSG]}         & $t_{GPGPU}$ (s) & 15.4 &  54.3 &  101.3 \\
\hline
Skylake/V100    & $t_{CPU}$ (s)   & 35.5 & 270.8 &  596.0 \\
{[PSG]}         & $t_{GPGPU}$ (s) & 11.6 &  47.7 &   76.5 \\
\hline
\end{tabular}
\caption{Comparison of total timings for iterations across different architectures for a 54-atom Si supercell (diamond structure).  All timings were taken from a single calculation.}
\label{tab:Si_iter_archs}
\end{table}

\begin{figure}
  \centering
  \includegraphics[trim={0cm 0cm 0cm 0cm},clip,angle=270,width=0.9\textwidth]{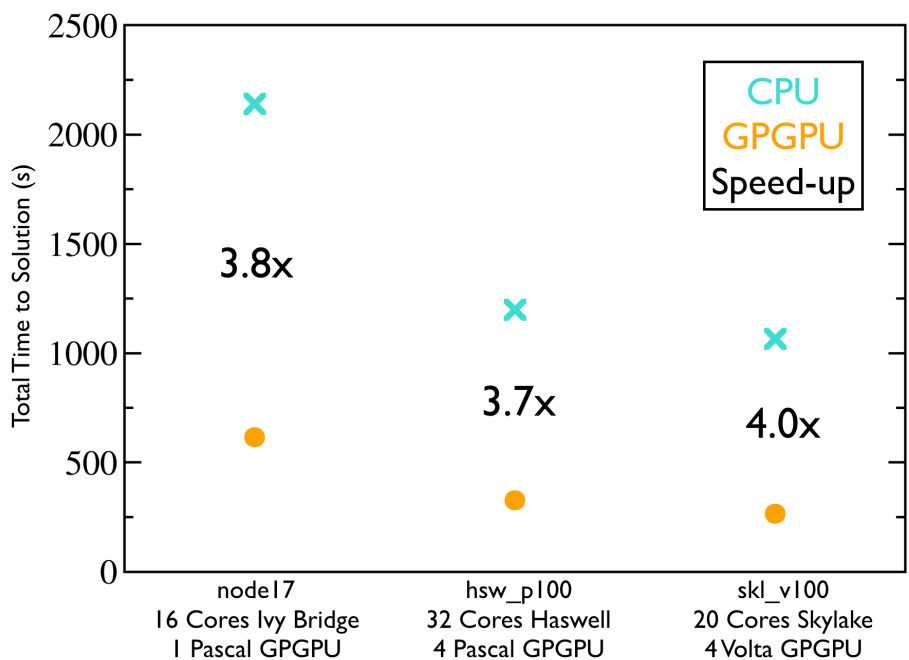}
    \caption{Overall timings across architectures for the calculation for a 54-atom Si supercell (diamond structure).  All calculations contain initialization, 12 SCF iterations, and 1 SCF + $\forceat$ + $\stresstens$ iteration.  Importantly, this also contains all non-GPU accelerated parts of the calculation}
  \label{fig:Si_overall_archs}
\end{figure}

While an individual SCF + $\forceat$ + $\stresstens$ iteration benefits greatly from GPGPU acceleration and takes considerably more time than a standard SCF iteration, it is calculated only when self-consistency is reached and often only needs to be calculated once (Fig. \ref{fig:SCF_Cycle}).  On the other hand, there will be multiple standard SCF iterations, which show a reduced GPGPU speedup.  To assess the impact of this difference in speedups on the overall timing of a calculation, Figure \ref{fig:Si_overall_archs} shows the timings for complete calculations for diamond Si on IvyBridge/GP100 [timewarp], Haswell/P100 [PSG], and Skylake/V100 [PSG] architectures.  Self-consistency is reached after 12 standard SCF iterations and one SCF + $\forceat$ + $\stresstens$ iteration, leading to speedups between 3.7 and 4.0 for the total time of the calculation.

We next turn our attention to the 103 material benchmark set.  To compare results from different materials, we estimate the expected workload for a given material using a metric
\begin{eqnarray}
  \#~\mathrm{ops} &=& \# B_{\nu} * MM \\
  MM &=& \langle |B_{\nu}| \rangle * \langle \mathrm{nnz}(B_{\nu}) \rangle^{2} \nonumber
\end{eqnarray}
where $\langle |B_{\nu}| \rangle$ is the average number of points per batch for a material,  $\langle \mathrm{nnz}(B_{\nu}) \rangle$ is the average number of basis functions with non-zero support per batch for a material, and $ \# B_{\nu} $ is the number of batches for a material.  The quantity $MM$ is then a measure of the timing for a hypothetical real-space operation involving a single matrix multiplication, and $\#~\mathrm{ops}$ is a measure of the total time of the real-space operation across all batches.  As motivated by Eq. (\ref{eq:batch_matrix_mult}), and as we show below, this metric will give an approximate estimate for the expected workload, as it is expected to map particularly the dominant part of the workload associated with dense serial matrix multiplications with high accuracy.

\begin{figure*}
  \centering
  \includegraphics[trim={2.4cm 0cm 2.3cm 0cm},clip,angle=270,width=\textwidth]{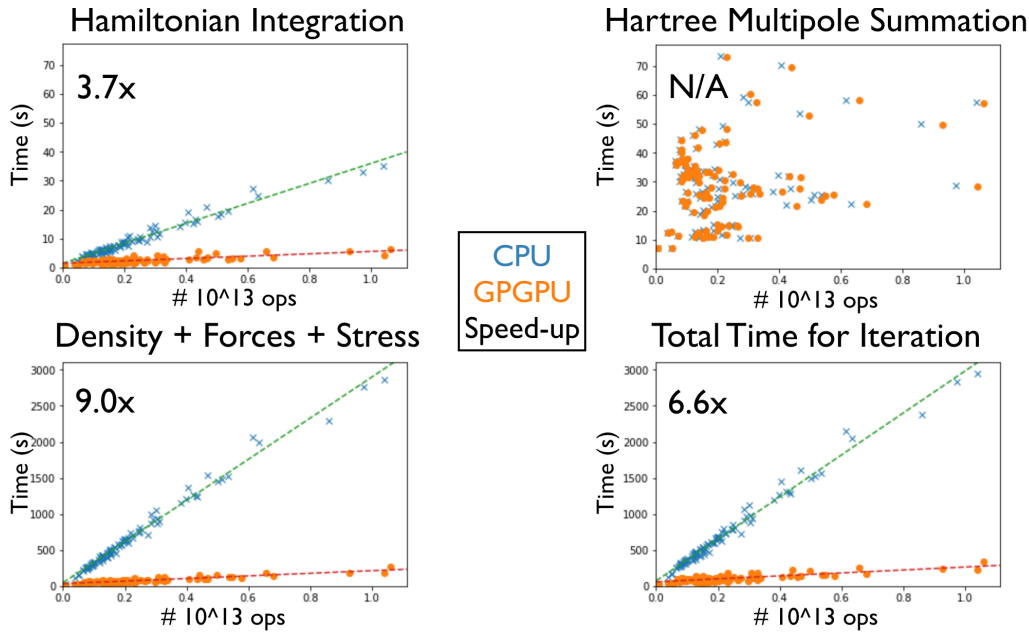}
    \caption{Timings for various operations in an SCF + forces + stress tensor iteration across the 103 material benchmark set for the (a) Hamiltonian integration, (b) Hartree multipole summation and Hellman-Feynman forces, (c) density and Pulay contributions to forces and stress tensor, culminating in (d) the total time for the iteration.  The Skylake/V100 [PSG] architecture was used.}
  \label{fig:103_mater_ops_skl_v100}
\end{figure*}

Shown in Figure \ref{fig:103_mater_ops_skl_v100} are timings for an SCF + $\forceat$ + $\stresstens$ iteration calculated using Skylake/V100 [PSG].  The proposed metric correlates well to the timings observed for real-space operations in FHI-aims.  A strong linear trend for the Hamiltonian integration (Fig. \ref{fig:103_mater_ops_skl_v100}a) and density + force + stress evaluation (Fig. \ref{fig:103_mater_ops_skl_v100}c) is observed, reflecting the relatively significant workload of dense linear algebra in these operations.  The total time for an iteration (Fig. \ref{fig:103_mater_ops_skl_v100}a) shows a similar trend, as it is dominated by the density + force + stress evaluation.  No such trend is observed in the Hartree multipole summation, which does not rely on dense linear algebra and is also not yet GPU accelerated in the present work.

Average speedups of 3.7, 9.0, and 6.6 are observed for Hamiltonian integration, density evaluation, and the total time for an SCF + $\forceat$ + $\stresstens$ iteration for the 103 material benchmark set.  These speedups are consistent with the speedups observed for diamond Si.  Table \ref{tab:103_mater_iter_archs} shows a comparison across architectures for the speedups calculated for the 103 material benchmark set.  These results are analogous to the results in Table \ref{tab:Si_iter_archs}, indicating that our analysis of trends observed in the diamond Si system extends to other materials and other architectures as well.

\begin{table}
\centering
\scriptsize
\begin{tabular}{|c|c|c|c|}
\hline
Architecture    & ~SCF~ & SCF          & SCF              \\
                &       & + $\forceat$ & + $\forceat$     \\
                &       &              & + $\stresstens$ \\    
\hline
\hline
IvyBridge/GP100 & 2.4 & 3.9 & 4.5 \\
{[timewarp]} & & & \\
\hline
Haswell/P100    & 2.1 & 4.5 & 6.6 \\
{[PSG]} & & & \\
\hline
Skylake/V100    & 2.4 & 6.6 & 6.6 \\
{[PSG]} & & & \\
\hline
\end{tabular}
\caption{Comparison of speedups for iterations across different architectures for the 103 material benchmark set.}
\label{tab:103_mater_iter_archs}
\end{table}

\subsection{Strong Scaling for CPU-Only and GPGPU-Accelerated Calculations}

\begin{figure*}
  \centering
  \includegraphics[trim={0cm 0cm 0cm 0cm}, clip, width=\textwidth]{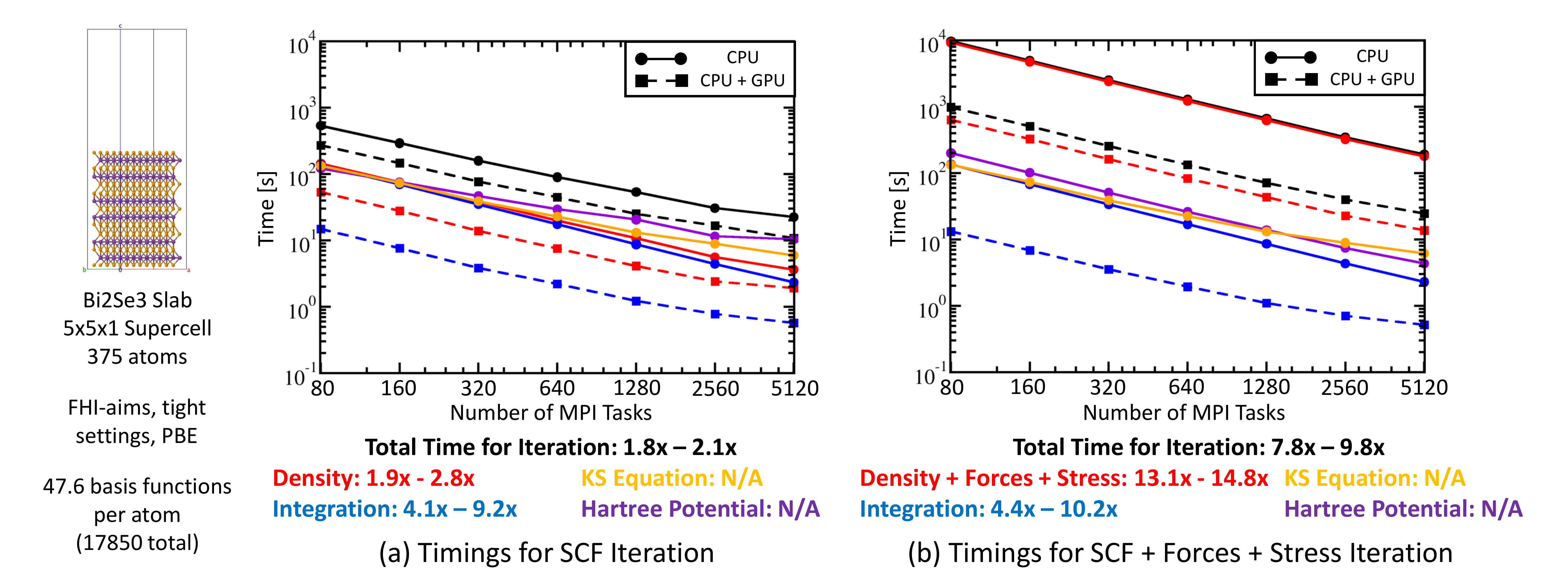}
    \caption{Timings on LLNL's Lassen for a 375-atom Bi$_2$Se$_3$ bilayer with a 40 \AA~vacuum for (a) an SCF iteration and (b) an SCF iteration including the calculation of forces and the stress tensor.}
  \label{fig:Bi2Se3_5x5x1}
\end{figure*}

The calculations in the previous section were performed using a fixed number of MPI tasks on a single node.  In the present section, we assess the scaling across nodes using a 375-atom \bise~bilayer system as a larger example. Strong scaling plots for the POWER9/V100 [LLNL] architecture are shown in Figure \ref{fig:Bi2Se3_5x5x1}, with subfigure (a) showing a standard SCF iteration and subfigure (b) showing an SCF + $\forceat$ + $\stresstens$ iteration.  Timings for 80, 160, 320, 640, 1280, 2560, and 5120 MPI tasks are presented, and we reintroduce the timings for the solution of the Kohn-Sham eigenvalue equation. The slab geometry corresponds to a (5$\times$5) supercell, using FHI-aims' ``tight'' production settings (17,850 basis functions total) and a $\Gamma$-point only calculation is performed. The calculations shown correspond to scalar-relativistic self-consistency iterations for this system, i.e., the workload that constitutes the bulk of a typical DFT calculation for such a system. As shown in~\cite{WPHuhn17} and others, the energy band structure of such a system is subject to spin-orbit coupling effects that would be essential to include at least at a post-processed second-variational (i.e., perturbative) level for qualitatively correct results. The necessary spin-orbit coupled Hamiltonian matrix integrals follow the exact same algorithms as those of the scalar-relativistic integrations shown in Fig.~\ref{fig:Bi2Se3_5x5x1}(a) and are therefore not repeated in detail in the plots. 

Timings for the \bise~bilayer are dominated by the real-space density update and potentially forces and stresses, consistent with the results from the previous section.  GPGPU-accelerated speedups of 1.8 - 2.1 and 7.8 - 9.8 are observed for overall timings in a standard SCF iteration and SCF + $\forceat$ + $\stresstens$ iterations, respectively.  Real-space operations show ideal scaling as a function of the number of MPI tasks, as is expected for the domain decomposition algorithm.  The solution of the Kohn-Sham matrix eigenvalue problem using the ELPA library~\cite{TAuckenthaler11,ELPA,PKus19_article,PKus19_book} (not GPU-accelerated in Fig.~\ref{fig:Bi2Se3_5x5x1}) shows reduced scaling due to the still relatively small matrix dimension size, but its diminished weight in these calculations does not meaningfully impact the overall scaling of the calculations. A denser $k$-space grid (e.g., 2$\times$2$\times$1 or 4$\times$4$\times$1) would increase the eigenvalue solver related workload and could be accelerated, e.g, by the recently published distributed-parallel, GPGPU-accelerated one-stage ELPA solver~\cite{PKus19_article,PKus19_book} (a computational step that is separate and independent of the real-space operations reported in this work).

\section{Conclusion}

In this paper, we show GPGPU acceleration for real-space operations relevant in semilocal DFT for production-quality materials simulations.  We particularly focus on the domain decomposition method used by the full-potential, all-electron FHI-aims electronic structure code for real-space operations.  We show that the time-intensive portions of the domain decomposition method are dense linear algebra operations which are bounded in memory consumption, allowing for efficient offloading to GPGPU resources.

The performance of the GPGPU acceleration in FHI-aims was assessed using a 103-material benchmark set on three heterogeneous CPU-GPGPU architectures.  GPGPU-accelerated speedups ranging from $s = 2.4$ for SCF iterations to $s = 6.6$ for SCF iterations including evaluation of forces $\forceat$ and the stress tensor $\stresstens$ were observed, with an overall estimated speedup of $s \approx 3-4$ expected for total times for entire calculations.  

Scaling on HPC resources was assessed on LLNL's Lassen calculations involving a 375-atom Bi$_2$Se$_3$ slab supercell.  The GPGPU-accelerated implementation shows near-ideal scaling similar to the CPU-only implementation.  We find that the evaluation of the density, forces, and stress tensors is the dominant computational workload for the 375-atom bilayer, allowing for overall GPGPU-accelerated speedups of s$\approx$8-10.  For significantly larger systems, the well-known cubic bottleneck of the Kohn-Sham DFT equation (the eigenvalue solver, which is not GPGPU-accelerated in this work) would begin to dominate the computational workload. However, a large subset of electronic structure computational needs in the range of small and mid-sized problems (below the range in which the eigenvalue solver dominates, i.e., up to several hundred atoms in this work) becomes accessible to GPGPU acceleration using the localized basis set strategies described above.

\section{Acknowledgements}

This work was supported by the LDRD Program of ORNL managed by UT-Battelle, LLC, for the U.S. DOE and by the Oak Ridge Leadership Computing Facility, which is a DOE Office of Science User Facility supported under Contract DE-AC05-00OR22725.  A portion of work was conducted at the Center for Nanophase Materials Sciences, which is a DOE Office of Science User Facility, and supported by the Creative Materials Discovery Program through the National Research Foundation of Korea funded by the Ministry of Science, ICT and Future Planning (NRF-2016M3D1A1919181). We gratefully acknowledge the support of NVIDIA Corporation with the donation of Quadro GP100 and Titan V GPGPUs used for local development, as well as access to their PSG cluster.  We thank Dr. Vincenzo Lordi and the Lawrence Livermore National Laboratory (LLNL), a U.S. Department of Energy Facility, for assistance with and access to LLNL's supercomputer Lassen for benchmarks conducted in this work.  The work on LLNL’s Lassen supercomputer was performed under the auspices of the U.S. Department of Energy at Lawrence Livermore National Laboratory under Contract No. DE-AC52-07NA27344.  We would finally like to acknowledge the contribution of Dr. Rainer Johanni, deceased in 2012, who pioneered the distributed-parallel CPU version of the locally-indexed real-space Hamiltonian scheme that is a critical foundation of this work.

\bibliographystyle{apsrev4-1}
\bibliography{wphbib}

\begin{thebibliography}{100}
\expandafter\ifx\csname url\endcsname\relax
  \def\url#1{\texttt{#1}}\fi
\expandafter\ifx\csname urlprefix\endcsname\relax\def\urlprefix{URL }\fi
\expandafter\ifx\csname href\endcsname\relax
  \def\href#1#2{#2} \def\path#1{#1}\fi

\bibitem{PHohenberg64}
P.~Hohenberg, W.~Kohn, Inhomogeneous electron gas, Phys. Rev. 136~(3B) (1964)
  864--871.
\newblock \href {http://dx.doi.org/10.1103/PhysRev.136.B864}
  {\path{doi:10.1103/PhysRev.136.B864}}.

\bibitem{WKohn65}
W.~Kohn, L.~J. Sham, Self-consistent equations including exchange and
  correlation effects, Phys. Rev. 140~(4A) (1965) 1133--1138.
\newblock \href {http://dx.doi.org/10.1103/PhysRev.140.A1133}
  {\path{doi:10.1103/PhysRev.140.A1133}}.

\bibitem{KBurke12}
K.~Burke, Perspective on density functional theory, J. Chem. Phys. 136 (2012)
  150901.
\newblock \href {http://dx.doi.org/10.1063/1.4704546}
  {\path{doi:10.1063/1.4704546}}.

\bibitem{PsiK14}
R.~O. Jones, Density functional theory: Past, present, ... future?, $\Psi_{k}$
  Newsletter 124 (2014) 1--23.

\bibitem{ADBecke14}
A.~D. Becke, Perspective: Fifty years of density-functional theory in chemical
  physics, J. Chem. Phys. 140 (2014) 18A301.
\newblock \href {http://dx.doi.org/10.1063/1.4869598}
  {\path{doi:10.1063/1.4869598}}.

\bibitem{AJain16}
A.~Jain, Y.~Shin, K.~A. Persson, Computational predictions of energy materials
  using density functional theory, Nature Rev. Mater. 1 (2016) 1--13.
\newblock \href {http://dx.doi.org/10.1038/natrevmats.2015.4}
  {\path{doi:10.1038/natrevmats.2015.4}}.

\bibitem{MJFrisch84}
M.~J. Frisch, J.~A. Pople, J.~S. Binkley, Self-consistent molecular orbital
  methods 25. {S}upplementary functions for {G}aussian basis sets, J. Chem.
  Phys. 80~(7) (1984) 3265--3269.
\newblock \href {http://dx.doi.org/10.1063/1.447079}
  {\path{doi:10.1063/1.447079}}.

\bibitem{THDunning89}
T.~H. Dunning, Gaussian basis sets for use in correlated molecular
  calculations. {I.} {T}he atoms boron through neon and hydrogen, J. Chem.
  Phys. 90~(2) (1989) 1007--1023.
\newblock \href {http://dx.doi.org/10.1063/1.456153}
  {\path{doi:10.1063/1.456153}}.

\bibitem{ASzabo96}
A.~Szabo, N.~S. Ostlund, Modern Quantum Chemistry: Introduction to Advanced
  Electronic Structure Theory, 1st Edition, Dover Publications, Mineola, NY,
  1996.

\bibitem{AKWilson96}
A.~K. Wilson, T.~{van Mourik}, T.~H. Dunning, Gaussian basis sets for use in
  correlated molecular calculations. {VI.} {S}extuple zeta correlation
  consistent basis sets for boron through neon, J. Mol. Struct. 388 (1996)
  339--349.
\newblock \href {http://dx.doi.org/10.1016/S0166-1280(96)80048-0}
  {\path{doi:10.1016/S0166-1280(96)80048-0}}.

\bibitem{FWeigend05}
F.~Weigend, R.~Ahlrichs, Balanced basis sets of split valence, triple zeta
  valence and quadruple zeta valence quality for {H} to {R}n: {D}esign and
  assessment of accuracy, Phys. Chem. Chem. Phys. 7 (2005) 3297--3305.
\newblock \href {http://dx.doi.org/10.1039/b508541a}
  {\path{doi:10.1039/b508541a}}.

\bibitem{MValiev10}
M.~Valiev, E.~J. Bylaska, N.~Govind, K.~Kowalski, T.~P. Straatsma, H.~J.~J.
  {Van Dam}, D.~Wang, J.~Nieplocha, E.~Apra, T.~L. Windus, W.~A. {de Jong},
  {NWChem}: A comprehensive and scalable open-source solution for large scale
  molecular simulations, Comput. Phys. Commun. 181 (2010) 1477--1489.
\newblock \href {http://dx.doi.org/10.1016/j.cpc.2010.04.018}
  {\path{doi:10.1016/j.cpc.2010.04.018}}.

\bibitem{JHutter14}
J.~Hutter, M.~Iannuzzi, F.~Schiffmann, J.~VandeVondele, cp2k: atomistic
  simulations of condensed matter systems, WIREs Comput. Mol. Sci. 4 (2014)
  15--25.
\newblock \href {http://dx.doi.org/10.1002/wcms.1159}
  {\path{doi:10.1002/wcms.1159}}.

\bibitem{FFurche14}
F.~Furche, R.~Ahlrichs, C.~H{\"a}ttig, W.~Klopper, M.~Sierka, F.~Weigend,
  Turbomole, WIREs Comput. Mol. Sci. 4 (2014) 91--100.
\newblock \href {http://dx.doi.org/10.1002/wcms.1162}
  {\path{doi:10.1002/wcms.1162}}.

\bibitem{YShao15}
Y.~Shao, Z.~Gan, E.~Epifanovsky, A.~T. Gilbert, M.~Wormit, J.~Kussmann, A.~W.
  Lange, A.~Behn, J.~Deng, X.~Feng, D.~Ghosh, M.~Goldey, P.~R. Horn, L.~D.
  Jacobson, I.~Kaliman, R.~Z. Khaliullin, T.~Ku{\'s}, A.~Landau, J.~Liu, E.~I.
  Proynov, Y.~M. Rhee, R.~M. Richard, M.~A. Rohrdanz, R.~P. Steele, E.~J.
  Sundstrom, H.~L. {Woodcock III}, P.~M. Zimmerman, D.~Zuev, B.~Albrecht,
  E.~Alguire, B.~Austin, G.~J.~O. Beran, Y.~A. Bernard, E.~Berquist,
  K.~Brandhorst, K.~B. Bravaya, S.~T. Brown, D.~Casanova, C.-M. Chang, Y.~Chen,
  S.~H. Chien, K.~D. Closser, D.~L. Crittenden, M.~Diedenhofen, R.~A. {DiStasio
  Jr.}, H.~Do, A.~D. Dutoi, R.~G. Edgar, S.~Fatehi, L.~Fusti-Molnar,
  A.~Ghysels, A.~Golubeva-Zadorozhnaya, J.~Gomes, M.~W. Hanson-Heine, P.~H.
  Harbach, A.~W. Hauser, E.~G. Hohenstein, Z.~C. Holden, T.-C. Jagau, H.~Ji,
  B.~Kaduk, K.~Khistyaev, J.~Kim, J.~Kim, R.~A. King, P.~Klunzinger,
  D.~Kosenkov, T.~Kowalczyk, C.~M. Krauter, K.~U. Lao, A.~D. Laurent, K.~V.
  Lawler, S.~V. Levchenko, C.~Y. Lin, F.~Liu, E.~Livshits, R.~C. Lochan,
  A.~Luenser, P.~Manohar, S.~F. Manzer, S.-P. Mao, N.~Mardirossian, A.~V.
  Marenich, S.~A. Maurer, N.~J. Mayhall, E.~Neuscamman, C.~M. Oana,
  R.~Olivares-Amaya, D.~P. O'Neill, J.~A. Parkhill, T.~M. Perrine, R.~Peverati,
  A.~Prociuk, D.~R. Rehn, E.~Rosta, N.~J. Russ, S.~M. Sharada, S.~Sharma, D.~W.
  Small, A.~Sodt, T.~Stein, D.~St{\"u}ck, Y.-C. Su, A.~J. Thom, T.~Tsuchimochi,
  V.~Vanovschi, L.~Vogt, O.~Vydrov, T.~Wang, M.~A. Watson, J.~Wenzel, A.~White,
  C.~F. Williams, J.~Yang, S.~Yeganeh, S.~R. Yost, Z.-Q. You, I.~Y. Zhang,
  X.~Zhang, Y.~Zhao, B.~R. Brooks, G.~K.~L. Chan, D.~M. Chipman, C.~J. Cramer,
  W.~A. {Goddard III}, M.~S. Gordon, W.~J. Hehre, A.~Klamt, H.~F. {Schaefer
  III}, M.~W. Schmidt, C.~D. Sherrill, D.~G. Truhlar, A.~Warshel, X.~Xu,
  A.~Aspuru-Guzik, R.~Baer, A.~T. Bell, N.~A. Besley, J.-D. Chai, A.~Dreuw,
  B.~D. Dunietz, T.~R. Furlani, S.~R. Gwaltney, C.-P. Hsu, Y.~Jung, J.~Kong,
  D.~S. Lambrecht, W.~Liang, C.~Ochsenfeld, V.~A. Rassolov, L.~V. Slipchenko,
  J.~E. Subotnik, T.~{Van Voorhis}, J.~M. Herbert, A.~I. Krylov, P.~M.~W. Gill,
  M.~Head-Gordon, Advances in molecular quantum chemistry contained in the
  {Q-Chem} 4 program package, Mol. Phys. 113~(2) (2015) 184--215.
\newblock \href {http://dx.doi.org/10.1080/00268976.2014.952696}
  {\path{doi:10.1080/00268976.2014.952696}}.

\bibitem{Gaussian16}
M.~J. Frisch, G.~W. Trucks, H.~B. Schlegel, G.~E. Scuseria, M.~A. Robb, J.~R.
  Cheeseman, G.~Scalmani, V.~Barone, G.~A. Petersson, H.~Nakatsuji, X.~Li,
  M.~Caricato, A.~V. Marenich, J.~Bloino, B.~G. Janesko, R.~Gomperts,
  B.~Mennucci, H.~P. Hratchian, J.~V. Ortiz, A.~F. Izmaylov, J.~L. Sonnenberg,
  D.~Williams-Young, F.~Ding, F.~Lipparini, F.~Egidi, J.~Goings, B.~Peng,
  A.~Petrone, T.~Henderson, D.~Ranasinghe, V.~G. Zakrzewski, J.~Gao, N.~Rega,
  G.~Zheng, W.~Liang, M.~Hada, M.~Ehara, K.~Toyota, R.~Fukuda, J.~Hasegawa,
  M.~Ishida, T.~Nakajima, Y.~Honda, O.~Kitao, H.~Nakai, T.~Vreven,
  K.~Throssell, J.~A. Montgomery, {Jr.}, J.~E. Peralta, F.~Ogliaro, M.~J.
  Bearpark, J.~J. Heyd, E.~N. Brothers, K.~N. Kudin, V.~N. Staroverov, T.~A.
  Keith, R.~Kobayashi, J.~Normand, K.~Raghavachari, A.~P. Rendell, J.~C.
  Burant, S.~S. Iyengar, J.~Tomasi, M.~Cossi, J.~M. Millam, M.~Klene, C.~Adamo,
  R.~Cammi, J.~W. Ochterski, R.~L. Martin, K.~Morokuma, O.~Farkas, J.~B.
  Foresman, D.~J. Fox, Gaussian 16 {R}evision {B}.01, {}Gaussian Inc.
  Wallingford CT (2016).

\bibitem{SRJensen17}
S.~R. Jensen, S.~Saha, J.~A. Flores-Livas, W.~Huhn, V.~Blum, S.~Goedecker,
  L.~Frediani, The elephant in the room of density functional theory
  calculations, J. Phys. Chem. Lett. 8 (2017) 1449--1457.
\newblock \href {http://dx.doi.org/10.1021/acs.jpclett.7b00255}
  {\path{doi:10.1021/acs.jpclett.7b00255}}.

\bibitem{JCSlater30}
J.~C. Slater, Atomic shielding constants, Phys. Rev. 36 (1930) 57--64.
\newblock \href {http://dx.doi.org/10.1103/PhysRev.36.57}
  {\path{doi:10.1103/PhysRev.36.57}}.

\bibitem{GteVelde01}
G.~te~Velde, F.~M. Bickelhaupt, E.~J. Baerends, C.~F. Guerra, S.~J.~A. van
  Gisbergen, J.~G. Snijders, T.~Ziegler, Chemistry with {ADF}, J. Comput. Chem.
  22~(9) (2001) 931--967.
\newblock \href {http://dx.doi.org/10.1002/jcc.1056}
  {\path{doi:10.1002/jcc.1056}}.

\bibitem{EvanLenthe03}
E.~van Lenthe, E.~J. Baerends, Optimized {S}later-type basis sets for the
  elements 1-118, J. Comput. Chem. 24~(9) (2003) 1142--1156.
\newblock \href {http://dx.doi.org/10.1002/jcc.10255}
  {\path{doi:10.1002/jcc.10255}}.

\bibitem{FWAverill73}
F.~W. Averill, D.~E. Ellis, An efficient numerical multicenter basis set for
  molecular orbital calculations: Application to {FeCl4}, J. Chem. Phys 59~(12)
  (1973) 6412--6418.
\newblock \href {http://dx.doi.org/10.1063/1.1680020}
  {\path{doi:10.1063/1.1680020}}.

\bibitem{AZunger77}
A.~Zunger, A.~J. Freeman, Self-consistent numerical-basis-set
  linear-combination-of-atomic-orbitals for the study of solids in the local
  density formalism, Phys. Rev. B 15~(10) (1977) 4716--4737.
\newblock \href {http://dx.doi.org/10.1103/PhysRevB.15.4716}
  {\path{doi:10.1103/PhysRevB.15.4716}}.

\bibitem{BDelley82}
B.~Delley, D.~E. Ellis, Efficient and accurate expansion methods for molecules
  in local density models, J. Chem. Phys 76~(4) (1982) 1949--1982.
\newblock \href {http://dx.doi.org/10.1063/1.443168}
  {\path{doi:10.1063/1.443168}}.

\bibitem{OFSankey89}
O.~F. Sankey, D.~J. Niklewski, Ab initio multicenter tight-binding model for
  molecular-dynamics simulations and other applications in covalent systems,
  Phys. Rev. B 40~(6) (1989) 3979--3995.
\newblock \href {http://dx.doi.org/10.1103/PhysRevB.40.3979}
  {\path{doi:10.1103/PhysRevB.40.3979}}.

\bibitem{BDelley90}
B.~Delley, An all-electron numerical method for solving the local density
  functional for polyatomic molecules, J. Chem. Phys 92~(1) (1990) 508--517.
\newblock \href {http://dx.doi.org/10.1063/1.458452}
  {\path{doi:10.1063/1.458452}}.

\bibitem{APHorsfield97}
A.~P. Horsfield, Efficient ab initio tight binding, Phys. Rev. B 56~(11) (1997)
  6594--6602.
\newblock \href {http://dx.doi.org/10.1103/PhysRevB.56.6594}
  {\path{doi:10.1103/PhysRevB.56.6594}}.

\bibitem{KKoepernik99}
K.~Koepernik, H.~Eschrig, Full-potential nonorthogonal local-orbital
  minimum-basis band-structure scheme, Phys. Rev. B 59 (1999) 1743--1757.
\newblock \href {http://dx.doi.org/10.1103/PhysRevB.59.1743}
  {\path{doi:10.1103/PhysRevB.59.1743}}.

\bibitem{JMSoler02}
J.~M. Soler, E.~Artacho, J.~D. Gale, A.~Garc{\'{i}}a, J.~Junquera,
  P.~Ordej{\'{o}}n, D.~S{\'{a}}nchez-Portal, The {SIESTA} method for ab-initio
  order-{N} materials simulation, J. Phys.: Condens. Matter 14 (2002)
  2745--2779.
\newblock \href {http://dx.doi.org/10.1088/0953-8984/14/11/302}
  {\path{doi:10.1088/0953-8984/14/11/302}}.

\bibitem{TOzaki05}
T.~Ozaki, H.~Kino, Efficient projector expansion for the ab initio {LCAO}
  method, Phys. Rev. B 72 (2005) 045121.
\newblock \href {http://dx.doi.org/10.1103/PhysRevB.72.045121}
  {\path{doi:10.1103/PhysRevB.72.045121}}.

\bibitem{Blum09}
V.~Blum, R.~Gehrke, F.~Hanke, P.~Havu, V.~Havu, X.~Ren, K.~Reuter,
  M.~Scheffler, Ab initio molecular simulations with numeric atom-centered
  orbitals, Comp. Phys. Comm. 180 (2009) 2175--2196.
\newblock \href {http://dx.doi.org/doi:10.1016/j.cpc.2009.06.022}
  {\path{doi:doi:10.1016/j.cpc.2009.06.022}}.

\bibitem{IYZhang13}
I.~Y. Zhang, X.~Ren, P.~Rinke, V.~Blum, M.~Scheffler, Numeric
  atom-centered-orbital basis sets with valence-correlation consistency from
  {H} to {Ar}, New. J. Phys. 15 (2013) 123033.
\newblock \href {http://dx.doi.org/10.1088/1367-2630/15/12/123033}
  {\path{doi:10.1088/1367-2630/15/12/123033}}.

\bibitem{JMPerezJorda95}
J.~M. P{\'e}rez-Jord{\'a}, W.~Yang, An algorithm for {3D} numerical integration
  that scales linearly with the size of the molecule, Chem. Phys. Lett. 241
  (1995) 469--476.
\newblock \href {http://dx.doi.org/10.1016/0009-2614(95)00665-Q}
  {\path{doi:10.1016/0009-2614(95)00665-Q}}.

\bibitem{REStratmann96}
R.~E. Stratmann, G.~E. Scuseria, M.~J. Frisch, Achieving linear-scaling in
  exchange-correlation density functional quadratures, Chem. Phys. Lett. 257
  (1996) 213--223.
\newblock \href {http://dx.doi.org/10.1016/0009-2614(96)00600-8}
  {\path{doi:10.1016/0009-2614(96)00600-8}}.

\bibitem{CFonsecaGuerra98}
C.~{Fonseca Guerra}, J.~G. Snijders, G.~te~Velde, E.~J. Baerends, Towards an
  order-{N} {DFT} method, Theor. Chem. Acc. 99~(6) (1998) 391--403.
\newblock \href {http://dx.doi.org/10.1007/s002140050353}
  {\path{doi:10.1007/s002140050353}}.

\bibitem{GEScuseria99}
G.~E. Scuseria, Linear scaling density functional calculations with {G}aussian
  orbitals, J. Phys. Chem. A 103~(25) (1999) 4782--4790.
\newblock \href {http://dx.doi.org/10.1021/jp990629s}
  {\path{doi:10.1021/jp990629s}}.

\bibitem{openMXManual}
T.~Ozaki, H.~Kino, J.~Yu, M.~Han, N.~Kobayashi, M.~Ohfuti, F.~Ishii, T.~Ohwaki,
  User's manual of OpenMX, http://www.openmx-square.org (2008).

\bibitem{Havu09}
V.~Havu, V.~Blum, P.~Havu, M.~Scheffler, Efficient {O(N)} integration for
  all-electron electronic structure calculation using numeric basis functions,
  J. Comput. Phys. 228 (2009) 8367--8379.
\newblock \href {http://dx.doi.org/doi:10.1016/j.jcp.2009.08.008}
  {\path{doi:doi:10.1016/j.jcp.2009.08.008}}.

\bibitem{ESLarsen01}
E.~S. Larsen, D.~McAllister, Fast matrix multiplies using graphics hardware,
  in: SC '01 Proceedings of the 2001 ACM/IEEE conference on Supercomputing, ACM
  Press, New York, New York, USA, 2001, p.~55.
\newblock \href {http://dx.doi.org/10.1145/582034.582089}
  {\path{doi:10.1145/582034.582089}}.

\bibitem{KMoreland03}
K.~Moreland, E.~Angel, The {FFT} on a {GPU}, in: HWWS '03 Proceedings of the
  ACM SIGGRAPH/EUROGRAPHICS conference on Graphics hardware, Eurographics
  Association, Aire-la-Ville, Switzerland, Switzerland, 2003, pp. 112--119.

\bibitem{ISUfimtsev08}
I.~S. Ufimtsev, T.~J. Mart{\'i}nez, Quantum chemistry on graphical processing
  units. 1. {S}trategies for two-electron integral evaluation, J. Chem. Theory
  Comput. 4 (2008) 222--231.
\newblock \href {http://dx.doi.org/10.1021/ct700268q}
  {\path{doi:10.1021/ct700268q}}.

\bibitem{ISUfimtsev09_scf}
I.~S. Ufimtsev, T.~J. Mart{\'i}nez, Quantum chemistry on graphical processing
  units. 2. {D}irect self-consistent-field implementation, J. Chem. Theory
  Comput. 5~(4) (2009) 1004--1015.
\newblock \href {http://dx.doi.org/10.1021/ct800526s}
  {\path{doi:10.1021/ct800526s}}.

\bibitem{ISUfimtsev09_scf_correction}
I.~S. Ufimtsev, T.~J. Mart{\'i}nez, Quantum chemistry on graphical processing
  units. 2. {D}irect self-consistent-field implementation, J. Chem. Theory
  Comput. 5~(11) (2009) 3138.
\newblock \href {http://dx.doi.org/10.1021/ct900433g}
  {\path{doi:10.1021/ct900433g}}.

\bibitem{ISUfimtsev09_grads}
I.~S. Ufimtsev, T.~J. Mart{\'i}nez, Quantum chemistry on graphical processing
  units. 3. {A}nalytical energy gradients, geometry optimization, and first
  principles molecular dynamics, J. Chem. Theory Comput. 5~(10) (2009)
  2619--2628.
\newblock \href {http://dx.doi.org/10.1021/ct9003004}
  {\path{doi:10.1021/ct9003004}}.

\bibitem{NLuehr16}
N.~Luehr, A.~Sisto, T.~J. Mart{\'{i }}nez, Gaussian basis set {Hartree-Fock},
  density functional theory, and beyond on {GPUs}, in: R.~C. Walker, A.~W.
  G{\"{o}}tz (Eds.), Electronic Structure Calculations on Graphics Processing
  Units: From Quantum Chemistry to Condensed Matter Physics, John Wiley \&
  Sons, Ltd., West Sussex, United Kingdom, 2016, Ch.~4, pp. 67--100.

\bibitem{KYasuda07}
K.~Yasuda, Two-electron integral evaluation on the graphics processor unit, J.
  Comput. Chem 29 (2007) 334--342.
\newblock \href {http://dx.doi.org/10.1002/jcc.20779}
  {\path{doi:10.1002/jcc.20779}}.

\bibitem{KYasuda08}
K.~Yasuda, Accelerating density functional calculations with graphics
  processing unit, J. Chem. Theory Comput. 4~(8) (2008) 1230--1236.
\newblock \href {http://dx.doi.org/10.1021/ct8001046}
  {\path{doi:10.1021/ct8001046}}.

\bibitem{XGonze16}
X.~Gonze, F.~Jollet, F.~{Abreu Araujo}, D.~Adams, B.~Amadon, T.~Applencourt,
  C.~Audouze, J.-M. Beuken, J.~Bieder, A.~Bokhanchuk, E.~Bousquet, F.~Bruneval,
  D.~Caliste, M.~C{\^{o}}t{\'{e}}, F.~Dahm, F.~{Da Pieve}, M.~Delaveau, M.~{Di
  Gennaro}, B.~Dorado, C.~Espejo, G.~Geneste, L.~Genovese, A.~Gerossier,
  M.~Giantomassi, Y.~Gillet, D.~R. Hamann, L.~He, G.~Jomard, J.~{Laflamme
  Janssen}, S.~{Le Roux}, A.~Levitt, A.~Lherbier, F.~Liu,
  I.~Luka{\v{c}}evi{\'{c}}, A.~Martin, C.~Martins, M.~J.~T. Oliveira,
  S.~Ponc{\'{e}}, Y.~Pouillon, T.~Rangel, G.-M. Rignanese, A.~H. Romero,
  B.~Rousseau, O.~Rubel, A.~A. Shukri, M.~Stankovski, M.~Torrent, M.~J. {Van
  Setten}, B.~{Van Troeye}, M.~J. Verstraete, D.~Waroquiers, J.~Wiktor, B.~Xu,
  A.~Zhou, J.~W. Zwanziger, Recent developments in the {ABINIT} software
  package, Comput. Phys. Commun. 205 (2016) 106 -- 131.
\newblock \href {http://dx.doi.org/10.1016/j.cpc.2016.04.003}
  {\path{doi:10.1016/j.cpc.2016.04.003}}.

\bibitem{HvanSchoot16}
H.~{van Schoot}, L.~Visscher, {GPU} acceleration for density functional theory
  with {S}later-type orbitals, in: R.~C. Walker, A.~W. G{\"{o}}tz (Eds.),
  Electronic Structure Calculations on Graphics Processing Units: From Quantum
  Chemistry to Condensed Matter Physics, John Wiley \& Sons, Ltd., West Sussex,
  United Kingdom, 2016, Ch.~5, pp. 101--114.

\bibitem{LGenovese09}
L.~Genovese, M.~Ospici, T.~Deutsch, J.-F. M{\'{e}}haut, A.~Neelov,
  S.~Goedecker, Density functional theory calculation on many-cores hybrid
  central processing unit-graphic processing unit architectures, J. Chem. Phys
  131 (2009) 034103.
\newblock \href {http://dx.doi.org/10.1063/1.3166140}
  {\path{doi:10.1063/1.3166140}}.

\bibitem{LGenovese16}
L.~Genovese, B.~Videau, D.~Caliste, J.-F. M{\'{e}}haut, S.~Goedecker,
  T.~Deutsch, Wavelet-based density functional theory on massively parallel
  hybrid architectures, in: R.~C. Walker, A.~W. G{\"{o}}tz (Eds.), Electronic
  Structure Calculations on Graphics Processing Units: From Quantum Chemistry
  to Condensed Matter Physics, John Wiley \& Sons, Ltd., West Sussex, United
  Kingdom, 2016, Ch.~6, pp. 115--134.

\bibitem{OSchutt16}
O.~Sch{\"{u}}tt, P.~Messmer, J.~Hutter, J.~VandeVondele, {GPU} accelerated
  sparse matrix-matrix multiplication for linear scaling density functional
  theory, in: R.~C. Walker, A.~W. G{\"{o}}tz (Eds.), Electronic Structure
  Calculations on Graphics Processing Units: From Quantum Chemistry to
  Condensed Matter Physics, John Wiley \& Sons, Ltd., West Sussex, United
  Kingdom, 2016, Ch.~8, pp. 173--190.

\bibitem{SHakala13}
S.~Hakala, V.~Havu, J.~Enkovaara, R.~Nieminen, Parallel electronic structure
  calculations using multiple graphics processing units ({GPU}s), in:
  P.~Manninen, P.~{\"{O}}ster (Eds.), Lecture Notes in Computer Science, Vol.
  7782, Springer-Verlag Berlin Heidelberg, 2013, pp. 63--76.

\bibitem{JYan13}
J.~Yan, L.~Li, C.~O'Grady, Graphics processing unit acceleration of the random
  phase approximation in the projector augmented wave method, Comput. Phys.
  Commun. 184~(12) (2013) 2728--2733.
\newblock \href {http://dx.doi.org/10.1016/j.cpc.2013.07.014}
  {\path{doi:10.1016/j.cpc.2013.07.014}}.

\bibitem{SHakala16}
S.~Hakala, J.~Enkovaara, V.~Havu, J.~Yan, L.~Li, C.~O'Grady, R.~M. Nieminen,
  Grid-based projector-augmented wave method, in: R.~C. Walker, A.~W.
  G{\"{o}}tz (Eds.), Electronic Structure Calculations on Graphics Processing
  Units: From Quantum Chemistry to Condensed Matter Physics, John Wiley \&
  Sons, Ltd., West Sussex, United Kingdom, 2016, Ch.~9, pp. 191--210.

\bibitem{WJia17}
W.~Jia, J.~Wang, X.~Chi, L.-W. Wang, {GPU} implementation of the linear scaling
  three dimensional fragment method for large scale electronic structure
  calculations, Comput. Phys. Commun. 211 (2017) 8--15.
\newblock \href {http://dx.doi.org/10.1016/j.cpc.2016.07.003}
  {\path{doi:10.1016/j.cpc.2016.07.003}}.

\bibitem{XAndrade12}
X.~Andrade, J.~Alberdi-Rodriguez, D.~A. Strubbe, M.~J.~T. Oliveira,
  F.~Nogueira, A.~Castro, J.~Muguerza, A.~Arruabarrena, S.~G. Louie,
  A.~Aspuru-Guzik, A.~Rubio, M.~A.~L. Marques, Time-dependent
  density-functional theory in massively parallel computer architectures: the
  {OCTOPUS} project, J. Phys.: Condens. Matter 24~(23) (2012) 233202.
\newblock \href {http://dx.doi.org/10.1088/0953-8984/24/23/233202}
  {\path{doi:10.1088/0953-8984/24/23/233202}}.

\bibitem{Xandrade13}
X.~Andrade, A.~Aspuru-Guzik, Real-space density functional theory on graphical
  processing units: Computational approach and comparison to {G}aussian basis
  set methods, J. Chem. Theory Comput. 9~(10) (2013) 4360--4373.
\newblock \href {http://dx.doi.org/10.1021/ct400520e}
  {\path{doi:10.1021/ct400520e}}.

\bibitem{XAndrade16}
X.~Andrade, A.~Aspuru-Guzik, Application of graphics processing units to
  accelerate real-space density functional theory and time-dependent density
  functional theory calculations, in: R.~C. Walker, A.~W. G{\"{o}}tz (Eds.),
  Electronic Structure Calculations on Graphics Processing Units: From Quantum
  Chemistry to Condensed Matter Physics, John Wiley \& Sons, Ltd., West Sussex,
  United Kingdom, 2016, Ch.~10, pp. 211--237.

\bibitem{KWilkinson13}
K.~Wilkinson, C.-K. Skylaris, Porting {ONETEP} to graphical processing
  unit-based coprocessors. {1.} {FFT} box operations, J. Comput. Chem. 34~(28)
  (2013) 2446--2459.
\newblock \href {http://dx.doi.org/10.1002/jcc.23410}
  {\path{doi:10.1002/jcc.23410}}.

\bibitem{LWang11}
L.~Wang, Y.~Wu, W.~Jia, W.~Gao, X.~Chi, L.-W. Wang, Large scale plane wave
  pseudopotential density functional theory calculations on {GPU} clusters, in:
  SC '11: Proceedings of 2011 International Conference for High Performance
  Computing, Networking, Storage and Analysis, ACM, New York, NY, USA, 2011,
  p.~71.
\newblock \href {http://dx.doi.org/10.1145/2063384.2063479}
  {\path{doi:10.1145/2063384.2063479}}.

\bibitem{WJia13_1}
W.~Jia, Z.~Cao, L.~Wang, J.~Fu, X.~Chi, W.~Gao, L.-W. Wang, The analysis of a
  plane wave pseudopotential density functional theory code on a {GPU} machine,
  Comput. Phys. Commun. 18~(1) (2013) 9--18.
\newblock \href {http://dx.doi.org/10.1016/j.cpc.2012.08.002}
  {\path{doi:10.1016/j.cpc.2012.08.002}}.

\bibitem{WJia13_2}
W.~Jia, J.~Fu, Z.~Cao, L.~Wang, X.~Chi, W.~Gao, L.-W. Wang, Fast plane wave
  density functional theory molecular dynamics calculations on multi-{GPU}
  machines, J. Comput. Phys. 251 (2013) 102--115.
\newblock \href {http://dx.doi.org/10.1016/j.jcp.2013.05.005}
  {\path{doi:10.1016/j.jcp.2013.05.005}}.

\bibitem{PWmat}
{PWmat}, http://www.pwmatus.com/ (accessed June 12, 2019).

\bibitem{LVogt08}
L.~Vogt, R.~Olivares-Amaya, S.~Kermes, Y.~Shao, C.~Amador-Bedolla,
  A.~Aspuru-Guzik, Accelerating resolution-of-the-identity second-order
  {M{\o}ller-Plesset} quantum chemistry calculations with graphical processing
  units, J. Phys. Chem. A 112~(10) (2008) 2049--2057.
\newblock \href {http://dx.doi.org/10.1021/jp0776762}
  {\path{doi:10.1021/jp0776762}}.

\bibitem{ROlivaresAmaya10}
R.~Olivares-Amaya, M.~A. Wilson, R.~G. Edgar, L.~Vogt, Y.~Shao,
  A.~Aspuru-Guzik, Accelerating correlated quantum chemistry calculations using
  graphical processing units and a mixed precision matrix multiplication
  library, J. Chem. Theory Comput. 6~(1) (2010) 135--144.
\newblock \href {http://dx.doi.org/10.1021/ct900543q}
  {\path{doi:10.1021/ct900543q}}.

\bibitem{FSpiga11}
F.~Spiga, I.~Girotto, {phiGEMM}: a {CPU}-{GPU} library for porting {Quantum
  ESPRESSO} on hybrid systems, in: PDP '12 Proceedings of the 2012 20th
  Euromicro International Conference on Parallel, Distributed and Network-based
  Processing, IEEE Computer Society, Washington, DC, USA, 2012, pp. 368--375.
\newblock \href {http://dx.doi.org/10.1109/PDP.2012.72}
  {\path{doi:10.1109/PDP.2012.72}}.

\bibitem{JRomero18}
J.~Romero, E.~Phillips, G.~Ruetsch, M.~Fatica, F.~Spiga, P.~Giannozzi, A
  performance study of {Quantum ESPRESSO’s} {PWscf} code on multi-core and
  {GPU} systems, in: S.~Jarvis, S.~Wright, S.~Hammond (Eds.), Lecture Notes in
  Computer Science, Vol. 10724, Springer International Publishing AG, 2018, pp.
  67--87.
\newblock \href {http://dx.doi.org/10.1007/978-3-319-72971-8\_4}
  {\path{doi:10.1007/978-3-319-72971-8\_4}}.

\bibitem{SMoore12}
S.~Moore, E.~Briggs, M.~Hodak, W.~Lu, J.~Bernholc, C.-W. Lee, Scaling the {RMG}
  quantum mechanics code, in: BW-XSEDE '12 Proceedings of the Extreme Scaling
  Workshop, University of Illinois at Urbana-Champaign Champaign, IL, USA,
  2012.

\bibitem{RMG}
{RMG} - a real space multigrid {DFT} code, http://www.rmgdft.org/ (accessed
  June 12, 2019).

\bibitem{SMaintz11}
S.~Maintz, B.~Eck, R.~Dronskowski, Speeding up plane-wave electronic-structure
  calculations using graphics-processing units, Comput. Phys. Commun. 182~(7)
  (2011) 1421--1427.
\newblock \href {http://dx.doi.org/10.1016/j.cpc.2011.03.010}
  {\path{doi:10.1016/j.cpc.2011.03.010}}.

\bibitem{MHutchinson12}
M.~Hutchinson, M.~Widom, {VASP} on a {GPU}: Application to exact-exchange
  calculations of the stability of elemental boron, Comput. Phys. Commun.
  183~(7) (2012) 1422--1426.
\newblock \href {http://dx.doi.org/10.1016/j.cpc.2012.02.017}
  {\path{doi:10.1016/j.cpc.2012.02.017}}.

\bibitem{MHacene12}
M.~Hacene, A.~Anciaux-Sedrakian, X.~Rozanska, D.~Klahr, T.~Guignon,
  P.~Fleurat-Lessard, Accelerating {VASP} electronic structure calculations
  using graphic processing units, J. Comput. Chem 33~(32) (2012) 2581--2589.
\newblock \href {http://dx.doi.org/10.1002/jcc.23096}
  {\path{doi:10.1002/jcc.23096}}.

\bibitem{MHutchinson16}
M.~Hutchinson, P.~Fleurat-Lessard, A.~Anciaux-Sedrakian, D.~Stosic,
  J.~B{\'{e}}́dorf, S.~Tariq, Plane-wave density functional theory, in: R.~C.
  Walker, A.~W. G{\"{o}}tz (Eds.), Electronic Structure Calculations on
  Graphics Processing Units: From Quantum Chemistry to Condensed Matter
  Physics, John Wiley \& Sons, Ltd., West Sussex, United Kingdom, 2016, Ch.~7,
  pp. 135--172.

\bibitem{cuBLAS}
{cuBLAS API Reference Guide}, http://docs.nvidia.com/cuda/cublas (accessed June
  12, 2019).

\bibitem{cuFFT}
{cuFFT API Reference Guide}, https://docs.nvidia.com/cuda/cufft/ (accessed June
  12, 2019).

\bibitem{Thrust}
{Thrust API Reference Guide}, https://docs.nvidia.com/cuda/thrust/ (accessed
  June 12, 2019).

\bibitem{ELPA}
A.~Marek, V.~Blum, R.~Johanni, V.~Havu, B.~Lang, T.~Auckenthaler, A.~Heinecke,
  H.-J. Bungartz, H.~Lederer, The {ELPA} library: scalable parallel eigenvalue
  solutions for electronic structure theory and computational science, J.
  Phys.: Condens. Matter 26 (2014) 213201.
\newblock \href {http://dx.doi.org/10.1088/0953-8984/26/21/213201}
  {\path{doi:10.1088/0953-8984/26/21/213201}}.

\bibitem{PKus19_book}
P.~K{\r{u}}s, A.~Marek, H.~Lederer, {GPU} optimization of large-scale
  eigenvalue solver, in: F.~A. Radu, K.~Kumar, I.~Berre, J.~M. Nordbotten,
  I.~S. Pop (Eds.), Lecture Notes in Computational Science and Engineering,
  Vol. 126, Springer Nature Switzerland AG, 2019, Ch.~9, pp. 123--131.
\newblock \href {http://dx.doi.org/10.1007/978-3-319-96415-7\_9}
  {\path{doi:10.1007/978-3-319-96415-7\_9}}.

\bibitem{PKus19_article}
P.~K{\r{u}}s, A.~Marek, S.~S. K{\"{o}}cher, H.-H. Kowalski, C.~Carbogno,
  C.~Scheurer, K.~Reuter, M.~Scheffler, H.~Lederer, Optimizations of the
  eigensolvers in the {ELPA} library, Parallel Comput. 85 (2019) 167--177.
\newblock \href {http://dx.doi.org/10.1016/j.parco.2019.04.003}
  {\path{doi:10.1016/j.parco.2019.04.003}}.

\bibitem{STomov10_article}
S.~Tomov, J.~Dongarra, M.~Baboulin, Towards dense linear algebra for hybrid
  {GPU} accelerated manycore systems, Parallel Comput. 36 (2010) 232--240.
\newblock \href {http://dx.doi.org/10.1016/j.parco.2009.12.005}
  {\path{doi:10.1016/j.parco.2009.12.005}}.

\bibitem{STomov10_conference}
S.~Tomov, R.~Nath, H.~Ltaief, J.~Dongarra, Dense linear algebra solvers for
  multicore with {GPU} accelerators, in: 2010 IEEE International Symposium on
  Parallel \& Distributed Processing, Workshops and Phd Forum (IPDPSW), IEEE
  Computer Society, 2010, pp. 1--8.
\newblock \href {http://dx.doi.org/10.1109/IPDPSW.2010.5470941}
  {\path{doi:10.1109/IPDPSW.2010.5470941}}.

\bibitem{JDongarra14}
J.~Dongarra, M.~Gates, A.~Haidar, J.~Kurzak, P.~Luszczek, S.~Tomov,
  I.~Yamazaki, Accelerating numerical dense linear algebra calculations with
  {GPUs}, in: V.~Kindratenko (Ed.), Numerical Computations with {GPUs},
  Springer International Publishing Switzerland, 2014, Ch.~1, pp. 3--28.
\newblock \href {http://dx.doi.org/10.1007/978-3-319-06548-9\_1}
  {\path{doi:10.1007/978-3-319-06548-9\_1}}.

\bibitem{FKnuth15}
F.~Knuth, C.~Carbogno, V.~Atalla, V.~Blum, M.~Scheffler, All-electron formalism
  for total energy strain derivatives and stress tensor components for numeric
  atom-centered orbitals, Comput. Phys. Commun. 190 (2015) 33--50.
\newblock \href {http://dx.doi.org/10.1016/j.cpc.2015.01.003}
  {\path{doi:10.1016/j.cpc.2015.01.003}}.

\bibitem{LNemec13}
L.~Nemec, V.~Blum, P.~Rinke, M.~Scheffler, Thermodynamic equilibrium conditions
  of graphene films on {SiC}, Phys. Rev. Lett. 111 (2013) 065502.
\newblock \href {http://dx.doi.org/10.1103/PhysRevLett.111.065502}
  {\path{doi:10.1103/PhysRevLett.111.065502}}.

\bibitem{SVLevchenko15}
S.~V. Levchenko, X.~Ren, J.~Wieferink, R.~Johanni, P.~Rinke, V.~Blum,
  M.~Scheffler, Hybrid functionals for large periodic systems in an
  all-electron, numeric atom-centered basis framework, Comput. Phys. Comm. 192
  (2015) 60--69.
\newblock \href {http://dx.doi.org/10.1016/j.cpc.2015.02.021}
  {\path{doi:10.1016/j.cpc.2015.02.021}}.

\bibitem{VWZYu18}
V.~W.-Z. Yu, F.~Corsetti, A.~Garc{\'i}a, W.~P. Huhn, M.~Jacquelin, W.~Jia,
  B.~Lange, J.~Lu, W.~Mi, A.~Seifitokaldani, {\'A}.~V{\'a}zquez-Mayagoitia,
  C.~Yang, H.~Yang, V.~Blum, {ELSI}: A unified software interface for
  {K}ohn-{S}ham electronic structure solvers, Comput. Phys. Commun. 222 (2018)
  267--285.
\newblock \href {http://dx.doi.org/10.1016/j.cpc.2017.09.007}
  {\path{doi:10.1016/j.cpc.2017.09.007}}.

\bibitem{lejaeghere2016reproducibility}
K.~Lejaeghere, G.~Bihlmayer, T.~Bj{\"o}rkman, P.~Blaha, S.~Bl{\"u}gel, V.~Blum,
  D.~Caliste, I.~E. Castelli, S.~J. Clark, A.~Dal~Corso, et~al.,
  Reproducibility in density functional theory calculations of solids, Science
  351~(6280) (2016) aad3000.
\newblock \href {http://dx.doi.org/10.1126/science.aad3000}
  {\path{doi:10.1126/science.aad3000}}.

\bibitem{XRen12}
X.~Ren, P.~Rinke, V.~Blum, J.~Wieferink, A.~Tkatchenko, A.~Sanfilippo,
  K.~Reuter, M.~Scheffler, Resolution-of-identity approach to {Hartree-Fock},
  hybrid density functionals, {RPA}, {MP2} and {GW} with numeric atom-centered
  orbital basis functions, New J. of Phys. 14 (2012) 053020.
\newblock \href {http://dx.doi.org/10.1088/1367-2630/14/5/053020}
  {\path{doi:10.1088/1367-2630/14/5/053020}}.

\bibitem{Ihrig2015}
A.~C. Ihrig, J.~Wieferink, I.~Y. Zhang, M.~Ropo, X.~Ren, P.~Rinke,
  M.~Scheffler, V.~Blum, Accurate localized resolution of identity approach for
  linear-scaling hybrid density functionals and for many-body perturbation
  theory, New J. of Phys. 17~(9) (2015) 093020.
\newblock \href {http://dx.doi.org/10.1088/1367-2630/17/9/093020}
  {\path{doi:10.1088/1367-2630/17/9/093020}}.

\bibitem{GW100}
M.~J. van Setten, F.~Caruso, S.~Sharifzadeh, X.~Ren, M.~Scheffler, F.~Liu,
  J.~Lischner, L.~Lin, J.~R. Deslippe, S.~G. Louie, C.~Yang, F.~Weigend, J.~B.
  Neaton, F.~Evers, P.~Rinke, {GW100}: Benchmarking {G0W0} for molecular
  systems, J. Chem. Theory Comput. 11~(12) (2015) 5665--5687.
\newblock \href {http://dx.doi.org/10.1021/acs.jctc.5b00453}
  {\path{doi:10.1021/acs.jctc.5b00453}}.

\bibitem{WPHuhn17}
W.~P. Huhn, V.~Blum, One-hundred-three compound band-structure benchmark of
  post-self-consistent spin-orbit coupling treatments in density functional
  theory, Phys. Rev. Mater. 1 (2017) 033803.
\newblock \href {http://dx.doi.org/10.1103/PhysRevMaterials.1.033803}
  {\path{doi:10.1103/PhysRevMaterials.1.033803}}.

\bibitem{TAuckenthaler11}
T.~Auckenthaler, V.~Blum, H.-J. Bungartz, T.~Huckle, R.~Johanni,
  L.~Kr{\"{a}}mer, B.~Lang, H.~Lederer, P.~Willems, Parallel solution of
  partial symmetric eigenvalue problems from electronic structure calculations,
  Parallel Comput. 37 (2011) 783--794.
\newblock \href {http://dx.doi.org/10.1016/j.parco.2011.05.002}
  {\path{doi:10.1016/j.parco.2011.05.002}}.

\bibitem{SGoedecker99}
S.~Goedecker, Linear scaling electronic structure methods, Rev. Mod. Phys.
  71~(4) (1999) 1085--1123.
\newblock \href {http://dx.doi.org/10.1103/RevModPhys.71.1085}
  {\path{doi:10.1103/RevModPhys.71.1085}}.

\bibitem{on_bowler_2012}
D.~R. Bowler, T.~Miyazaki, {O(N)} methods in electronic structure calculations,
  Reports on Progress in Physics 75~(3) (2012) 036503.
\newblock \href {http://dx.doi.org/10.1088/0034-4885/75/3/036503}
  {\path{doi:10.1088/0034-4885/75/3/036503}}.

\bibitem{pexsi_lin_2013}
L.~Lin, M.~Chen, C.~Yang, L.~He, Accelerating atomic orbital-based electronic
  structure calculation via pole expansion and selected inversion, Journal of
  Physics: Condensed Matter 25~(29) (2013) 295501.
\newblock \href {http://dx.doi.org/10.1088/0953-8984/25/29/295501}
  {\path{doi:10.1088/0953-8984/25/29/295501}}.

\bibitem{VILebedev75}
V.~I. Lebedev, Values of the nodes and weights of quadrature formulas of
  {Gauss-Markov} type for a sphere from the ninth to seventeenth order of
  accuracy that are invariant with respect to an octahedron group with
  inversion, Zh. Vychisl. Mat. Mat. Fiz. 15~(1) (1975) 48--54.
\newblock \href {http://dx.doi.org/10.1016/0041-5553(75)90133-0}
  {\path{doi:10.1016/0041-5553(75)90133-0}}.

\bibitem{VILebedev76}
V.~I. Lebedev, Quadratures on a sphere, Zh. Vychisl. Mat. Mat. Fiz. 16~(2)
  (1976) 293--306.
\newblock \href {http://dx.doi.org/10.1016/0041-5553(76)90100-2}
  {\path{doi:10.1016/0041-5553(76)90100-2}}.

\bibitem{VILebedev99}
V.~I. Lebedev, D.~N. Laikov, A quadrature formula for the sphere of the 131st
  algebraic order of accuracy, Dokl. Math. 59 (1999) 477--481.

\bibitem{BDelley96}
B.~Delley, High order integration schemes on the unit sphere, J. Comput. Chem.
  17~(9) (1996) 1152--1155.

\bibitem{ADBecke88}
A.~D. Becke, A multicenter numerical integration scheme for polyatomic
  molecules, J. Chem. Phys. 88~(4) (1988) 2547--2553.
\newblock \href {http://dx.doi.org/10.1063/1.454033}
  {\path{doi:10.1063/1.454033}}.

\bibitem{PerdewBurkeErnzerhof96}
J.~P. Perdew, K.~Burke, M.~Ernzerhof, {G}eneralized {G}radient {A}pproximation
  made simple, Phys. Rev. Lett. 77~(18) (1996) 3865--3868.
\newblock \href {http://dx.doi.org/10.1103/PhysRevLett.77.3865}
  {\path{doi:10.1103/PhysRevLett.77.3865}}.

\bibitem{BDelley96_es}
B.~Delley, Fast calculation of electrostatics in crystals and large molecules,
  J. Phys. Chem. 100~(15) (1996) 6107--6110.
\newblock \href {http://dx.doi.org/10.1021/jp952713n}
  {\path{doi:10.1021/jp952713n}}.

\end{thebibliography}

\end{document}